\documentclass{aa} 

\usepackage{graphicx}
\usepackage{txfonts}
\usepackage{natbib}
\usepackage{capt-of}
\usepackage{subfigure}
\usepackage{soul}
\newcommand{\orcid}[1]{\unskip\protect\href{https://orcid.org/#1}{\protect\includegraphics[width=8pt,clip]{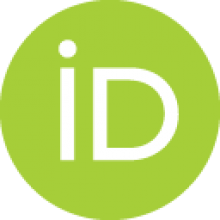}}}
\usepackage[colorlinks=true,allcolors=blue]{hyperref}
\makeatletter
\renewcommand*\aa@pageof{, page \thepage{} of \pageref*{LastPage}}
\makeatother
\begin{document}

   \title{Power of wavelets in analyses of transit and phase curves in
the presence of stellar variability and instrumental noise}

   \subtitle{II. Accuracy of the transit parameters}

   \author{Sz. Kálmán
          \inst{1,2,3,4}\orcid{0000-0003-3754-7889}
          \and
          Gy. M. Szabó \inst{1,5,6}\orcid{0000-0002-0606-7930}
          \and
          Sz. Csizmadia \inst{7}\orcid{0000-0001-6803-9698}
          }

   \institute{ MTA-ELTE Exoplanet Research Group, H-9700, Szent Imre h. u. 112, Szombathely, Hungary              
         \and
            ELTE Eötvös Loránd University, Doctoral School of Physics, Budapest, H-1117 Budapest Pázmány Péter sétány 1/A, Hungary
        \and     
            Konkoly Observatory, Research Centre for Astronomy and Earth Sciences,
            Eötvös Loránd Research Network (ELKH), Konkoly-Thege Miklós út 15–17, H-1121 Budapest, Hungary
         \and
         CSFK, MTA Centre of Excellence, Budapest, Konkoly Thege Miklós út 15-17., H-1121, Hungary\\
            \email{xilard1@gothard.hu}            
        \and
            ELTE Eötvös Loránd University, Gothard Astrophysical Observatory, H-9700, Szent
            Imre h. u. 112, Szombathely, Hungary
        \and
            MTA-ELTE Lendület "Momentum" Milky Way Research Group, Szent Imre h. u. 112, H-9700, Szombathely, Hungary
        \and 
            Deutsches Zentrum für Luft- und Raumfahrt, Institute of Planetary Research, Rutherfordtstrasse 2, D-12489 Berlin, Germany}

   \date{Recieved ....; accepted ....}

 
  \abstract
   {Correlated noise in exoplanet light curves, such as noise from stellar activity, convection noise, and instrumental noise, distorts the exoplanet transit light curves and leads to biases in the best-fit transit parameters. An optimal fitting algorithm can provide stability against the presence of correlated noises and lead to statistically consistent results, namely, the actual biases are usually within the error interval. This is not automatically satisfied by most of the algorithms in everyday use and the testing of the algorithms is necessary.}
   {In this paper, we describe a bootstrapping-like test to handle with the general case and we apply it to the wavelet-based  Transit and Light Curve Modeller (TLCM) algorithm, testing it for the stability against the correlated noise. We compare and contrast the results with regard to the FITSH algorithm, which is based on an assumption of white noise .}
   {We simulated transit light curves with previously known parameters in the presence of a correlated noise model generated by an  Autoregressive Integrated Moving Average (ARIMA) process. Then we solved the simulated observations and examined the resulting parameters and error intervals.}
   {We have found that the assumption of FITSH, namely, that only white noise is present, has led to inconsistencies in the results:\ the distribution of best-fit parameters is then broader than the determined error intervals by a factor of 3-6. On the other hand, the wavelet-based TLCM algorithm handles the correlated noise properly, leading to both properly determined parameter and error intervals that are perfectly consistent with the actual biases.}
   {}

\titlerunning{Accuracy of transit parameters}
   \maketitle
%


\section{Introduction}

The precise and accurate determination of parameters that characterise an exoplanet and its orbit are of key importance. Stellar activity such as spots (see e.g. \citealt{2010A&A...512A..38L,2013MNRAS.430.3032B,2014MNRAS.443.2517H,Mazeh2015}) or pulsation (\citealt{2014A&A...561A..48V,2018MNRAS.481.2871S}), instrumental effects \citep{2013ApJS..208...16M}, and even binning \citep{2010MNRAS.408.1758K, 2017Ap&SS.362..112J} result in the appearance of correlated noise in the studied light curves. As a consequence, the smooth light curve distortions, jumps, and sudden flux-changes lead to biases in the best-fit parameters, and the error intervals may be also widened. In case of small distortions, the uncertainty intervals will be still consistent with the real values of parameters. The biases are more egregious if the best-fit parameters end up more biased than their derived uncertainty range at a given high confidence level, leading to possible internal inconsistencies in the derived parameters and inaccuracies in the derived datasets, and suggest, for instance, transit timing variations (TTV) which have no physical origin.
The parameter biases, along with the question of whether the resulting parameter$+$error pairs are consistent, should be evaluated for each given case (transit and noise parameters and sampling), unless the applied algorithm is tested prudently.

In the following, we give an example of the inconsistencies and inaccuracies introduced by the time-correlated noise using the case of TTVs. These variations may be caused by a perturbing object in the system such as a stellar mass companion of the host or another planet\citep{2005Sci...307.1288H, 2005MNRAS.359..567A, 2011A&A...528A..53B}, by exomoons \citep{2007A&A...470..727S}, stellar activity (see e.g. \citealt{2013A&A...549A..35O}), or other non-dynamical phenomena \citep{ 2013A&A...553A..17S}. Since TTVs have a wide range of possible origins, their analysis may be applied of many different phenomena in a given exoplanetary system. We argue, however, that correlated noise may cause phenomena that mimic TTVs, making proper noise handling essential as underestimating the uncertainties may result in false TTV detections, while overestimating them may hide real phenomena.

The amount of photometric data from exoplanet systems is continuously increasing. This can primarily be attributed to the high-precision photometry of space missions such as \verb|CoRoT| \citep{2006cosp...36.3749B}, \verb|Kepler| \citep{2010Sci...327..977B}, \verb|K2| \citep{2014PASP..126..398H}, \verb|TESS| \citep{2015JATIS...1a4003R}, \verb|CHEOPS| \citep{2021ExA....51..109B}, \verb|PLATO| \citep{2018SPIE10698E..4XM}, and \verb|Ariel| \citep{2021arXiv210404824T}. The increasing amount of data requires more and more automatic data analysis and parameter determination. As we will be comparing data from different space- and ground-based telescopes, evaluated by different algorithms, consistency issues shall arise. The various research teams also have sometimes different statistical standards, which makes the comparison of results and accuracies even more difficult. 

There are a number of software that are used for the modelling of transit light curves (and/or RV data) including \textsc{EXOFAST} \citep{2013PASP..125...83E}, \textsc{pycheops}\footnote{https:\/github.com/pmaxted/pycheops}, \textsc{pyaneti} \citep{pyaneti2019}, \textsc{PyTransit} \citep{2015MNRAS.450.3233P}, and \textsc{TLCM}\citep{2020MNRAS.496.4442C}. 

These algorithms follow two general approaches. They either include a white noise assumption (i.e. assuming that the noise is completely uncorrelated) or a reconstruction of the noise correlations, 
and, in some of them, there is a possibility to switch between the two approaches within one code. The most widespread method to handle the correlated noise is to reconstruct it as a Gaussian process (GP) \citep{2017AJ....154..220F} or taking a wavelet approach \citep{2009ApJ...704...51C}. These two approaches (GP and wavelet) mostly differ in the fact that in GP there is a normality assumption in the noise (see e.g. \citealt{10.1214/aoms/1177730391}), whereas in the wavelet-based method there is no normality assumption and its inner parametrisation can (more or less force) the solution to be close to the priors.


In this paper, we test the stability, precision, and accuracy of parameter determination with wavelet-based methods using the Transit and Light Curve Modeller (TLCM; \citealt{2020MNRAS.496.4442C}) algorithm. For the sake of comparison, a linear optimisation approach with pattern determination was used via \verb|lfit|, the transit analyzing software from the FITSH software package  \citep{2012MNRAS.421.1825P}.

In \citet[hereafter Paper I]{2021arXiv210811822C}, a regularisation condition was added in the form of a Bayesian-prior to the wavelet-analysis technique. This prior compares the mean residuals of the fit to the mean photometric uncertainties of the observations to avoid overfitting. Paper I validated this approach by taking 310 ten-day long segments of Kepler Q1 SAP light curves. These light curves exhibited all common noise sources: spot activity, spot decay, flares, microflares, granulation, and pulsation, as well as cosmic ray hits, sudden flux changes up and down (flux-jumps), data leaks, and so on. Data leaks are gaps (series of missing data) in the continuity of the photometric time series. Such gaps can be planned, as in the case of TESS, where the data are downloaded from the satellite for typically 1.5 days in the middle of every sector and there are no observations carried out during it. They can be unplanned, for example, in the case of the safe-mode operation of a telescope or deleted data due to any kind of flag, caveat, or operation failure). Simulated transits, occultations, and phase curves were injected into these light curves and every light curve had five such realisations of synthetic data injections. In total, the retrieval of the parameters were performed on 1550 realisations and conclusions were drawn from the comparison of the injected and retrieved parameters. No pre-filtering of the light curves were made before modelling, the transit$+$occultation$+$phase curves (including reflection, beaming, and ellipsoidal effects) model were fitted simultaneously with the wavelet-based noise model. Paper I determined the minimum signal-to-noise ratios needed to retrieve the individual parameters.

\citet[submitted, hereafter Paper III]{kelt9pow} contains an application of the technique to KELT-9b. More accurate transit parameters were obtained by reducing the effect of correlated noise in the light curves than in previous studies reported.

This paper is structured as follows. In Section \ref{sec:meth}, we describe the simulating process, including the time-correlated noise model and the software used for light curve analysis. In Section \ref{sec:res}, we present an approach for error estimation without treating the correlated noise, while showing the biasing that the red noise induces in the transit parameters, as well as presenting the power of wavelet formulation in handling the time-correlated noise. We also present a possible application in TTV analysis. We give our conclusions in Section \ref{sec:sum}.


\section{Methods} \label{sec:meth}

In order to examine the biasing induced by the red noise, we made use of a synthetic light curve based on the model of \cite{2002ApJ...580L.171M} and simulated using the FITSH/\textit{lfit} software \citep{2012MNRAS.421.1825P}. This code describes the transit light curve of an exoplanet in terms of the time of the midstransit, $t_C$, the ratio of the planetary and stellar radii, $p=R_P/R_S$, the squared impact parameter, $b^2=R_S/a \cos i$, and the $\omega= (a/R_S) (2 \pi / P) (1-b^2)^{-1/2}$ parameter \citep{2009PhDT.........2P}, where $P$ is the orbital period, $i$ is the inclination of the orbital plane relative to the life of sight, and $a$ is the semi-major axis. The limb darkening of the host star was taken into account using a quadratic formula with the parameters $u_1$ and $u_2$. We assumed a circular orbit and a spherically symmetric star and planet. The values chosen for each parameter (Table \ref{table:params}), correspond roughly to a hot Jupiter but are otherwise arbitrary. 

\begin{table}
\caption{Parameters for the simulated light curve.} 
\label{table:params}      
\centering 
\begin{tabular}{c c c c c}        
\hline 
\hline
$P$ [d] & $t_C$ [HJD] & $p$ & $b^2$ & $\omega$ [d$^{-1}$] \\    
\hline           
   1.0914222 & 2455229.31035 & 0.1416 & 0.460 & 19.328 \\ 
\hline                              
\end{tabular}
\end{table}


We added a randomly selected segment of a simulated correlated noise model to the synthetic light curve, thereby constructing a realistic transit light curve which we then solved using the packages FITSH/\textit{lfit} and TLCM. We repeated this process until we had a statistically significant sample size that enabled the tests for parameter stability. For the sake of comparison, we repeated this process using only white noise and the FITSH/\textit{lfit} code. This approach is similar to naive bootstrapping, however, instead of simply resampling the noise model, as is typically done in the classical non-parametric bootstrapping  step, we replaced the entire selected realisation of the partly correlated noise. Our approach also resembles a Monte Carlo (MC) simulation in the sense of simulating a measurement with noise, however, in the case of the MC, the parameter values are sought along the parameter grid; here, we are mapping the error interval of the input parameters from a reconstruction.

\subsection{Noise model}

The spectral density of correlated noises is $\propto 1/f^\gamma$, where $f$ is the frequency, with $\gamma=0$ being the special case of 'white noise' (i.e. totally uncorrelated noise). There are various terms used for the correlated noises, and in some works, it is only the $\gamma = 2$ case that is called 'red noise', however, other scenarios are also commonly referred to as 'red', including the $\gamma=1$ case of 'pink noise'. The noise model used in this study is characterised by $\gamma \approx 1.2$, thus we shall refer to it as correlated noise or red noise.

\begin{figure*}
    \centering
    \includegraphics[width = \textwidth]{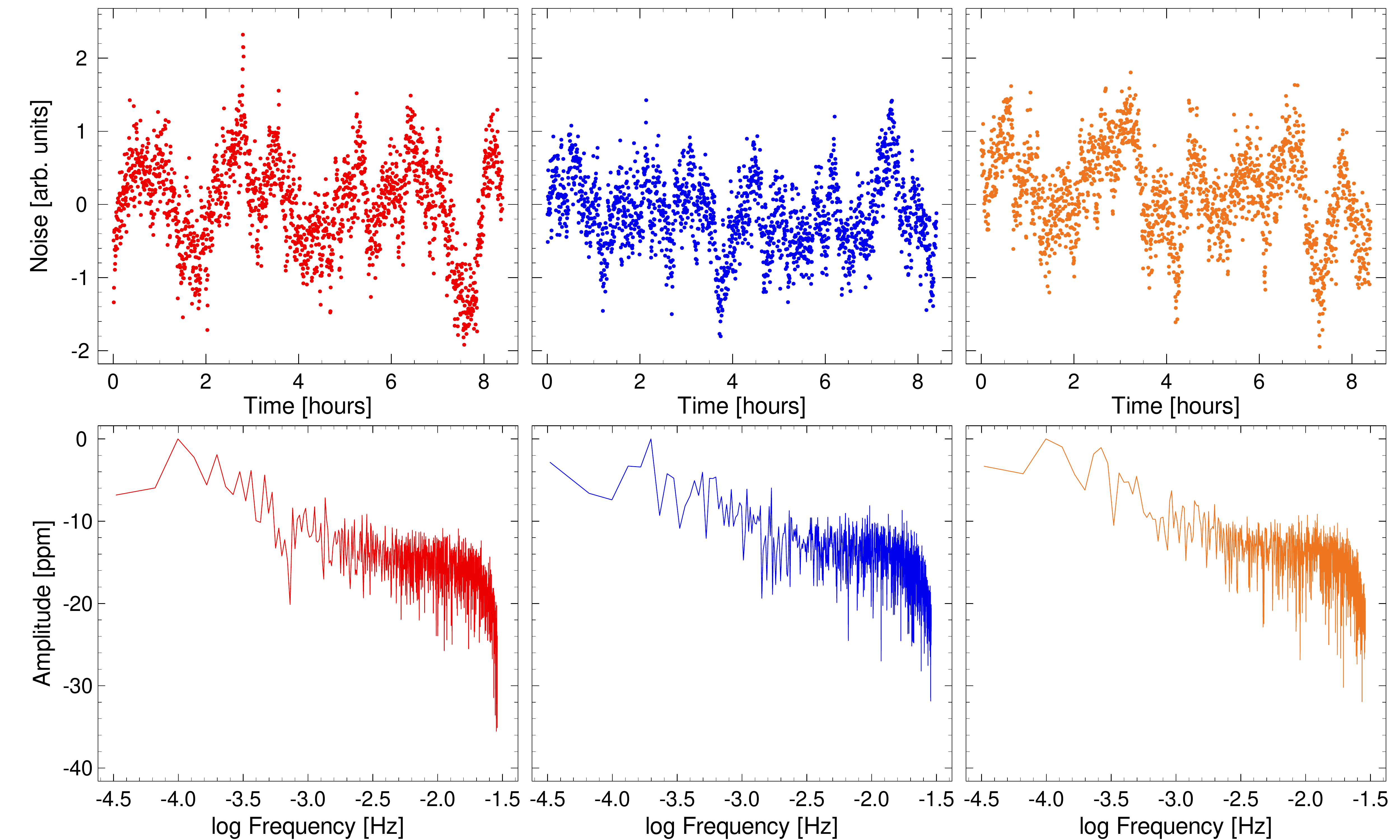}
    \caption{Three different randomly selected realisations of the noise model (top row), with durations equal to that of the input light curve, and their spectra (bottom row).}
    \label{fig:noise}
\end{figure*}

In order to get a sufficiently long noise model that retains its characteristics, yet still differs from realisation to realisation, we cloned our model via an ARIMA\footnote{Autoregressive Integrated Moving Average} process (as input, we used the publicly available data of \cite{2018SciA....4.1784T}) to clone an HST-like noise. Different segments and their respective spectra are shown in Fig. \ref{fig:noise}. Just by looking at Fig. \ref{fig:noise}, we can easily deduce that our model is time-correlated and quasi-periodic; these properties result in the distortion of the transit parameters. Converting into the magnitude system used during the fitting, the red noise model has a mean value of $\approx 6.3$ mmag and a standard deviation of $\approx 586.9$ mmag.

\subsection{Solving the light curve}
The simulated noisy transit light curves were solved using two different software programmes: FITSH/\textit{lfit} and TLCM. Here, we present a brief description of the relevant fitting processes. 

\begin{figure}
   \centering
   \includegraphics[width=\hsize]{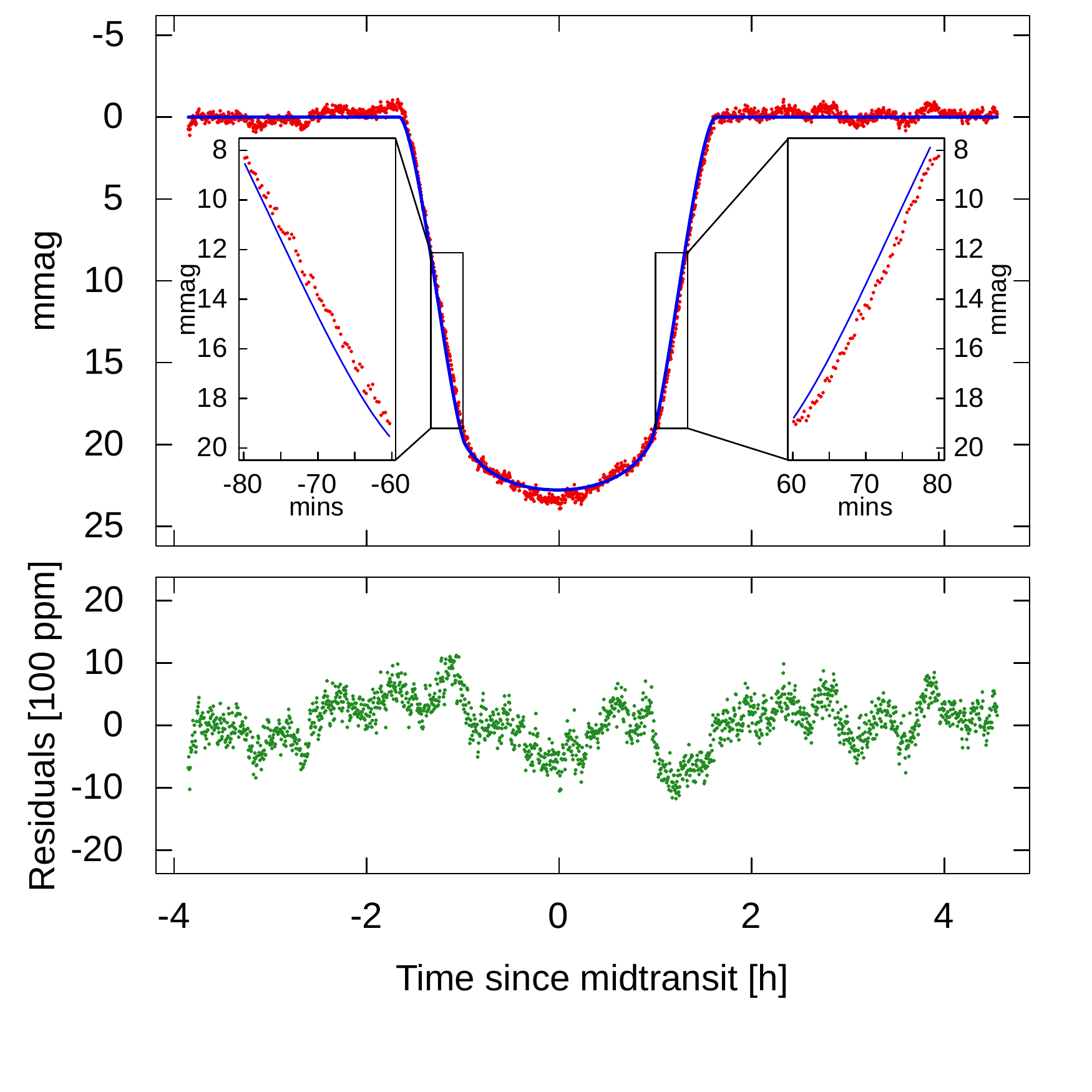}
      \caption{Typical example of a simulated noise transit light curve (upper panel, red), its \textit{lfit} solution (upper panel, blue) and the residuals (lower panel). Note: both during the ingress an egress phases, the blue curve is shifted to the left of the simulated points, meaning that purely the presence of correlated noise can induce TTV-like phenomena. The difference between the simulated and best-fit curves during these phases is emphasised using the two inset plots.}
         \label{fig:typic}
   \end{figure}

The first code, \textit{lfit}, uses an extended Markov chain Monte-Carlo (XMMC) algorithm to minimise $\chi^2$ during the fitting process, while the error estimation is done through refitting to synthetic data sets (EMCE) and XMMC \citep{2009PhDT.........2P}. Although it accomplishes the fitting process reasonably quickly, it does not account for the correlated noise. We treated $t_C$, $p$, $b^2$, and $\omega$ as free parameters, while $P$, $u_1$, and $u_2$ were set to the input values. A typical light curve is shown in Fig. \ref{fig:typic}. In this study, we  fit 10,000 light curves using this software. 

\begin{figure*}
    \centering
    \includegraphics[width=0.45\hsize]{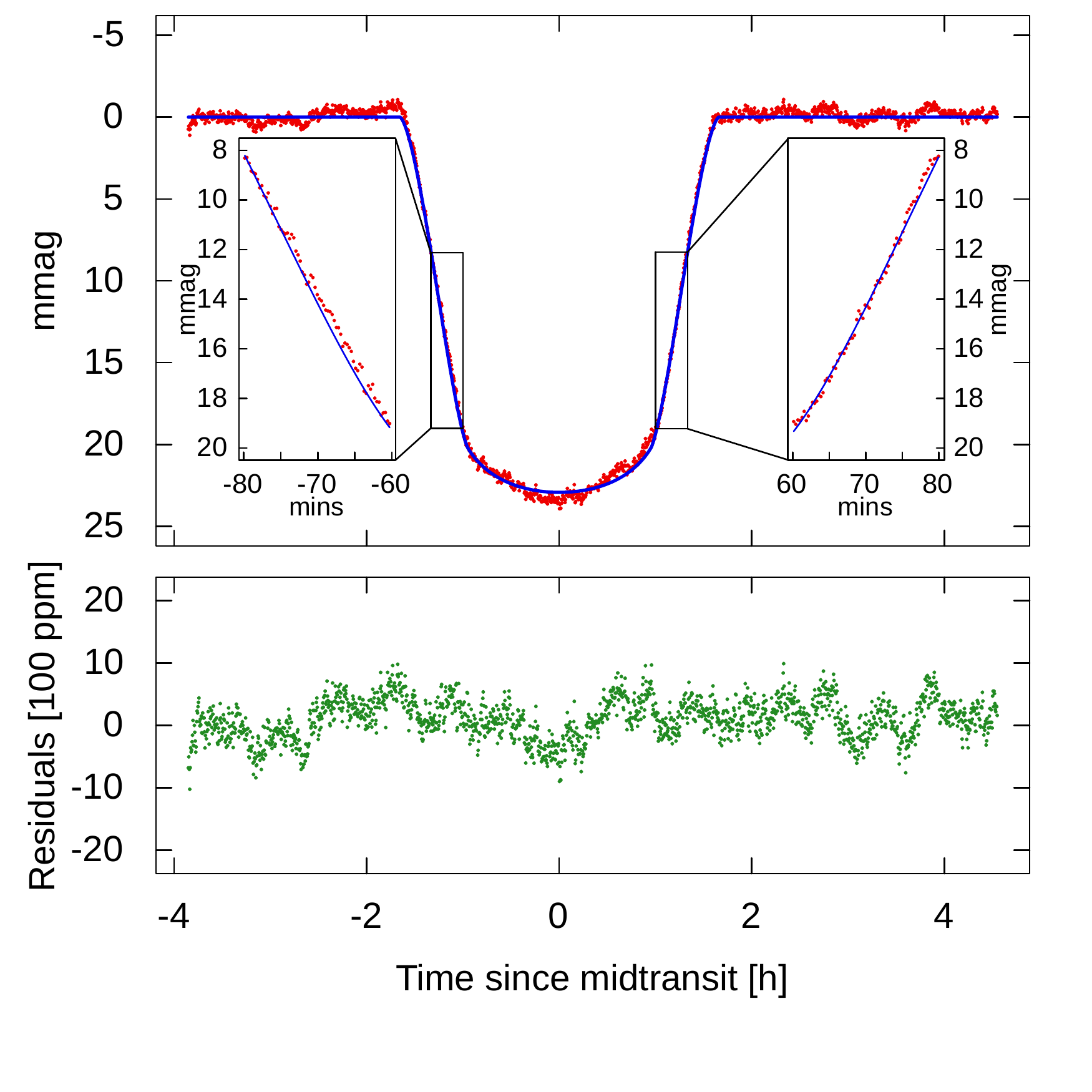}
    \includegraphics[width=0.45\hsize]{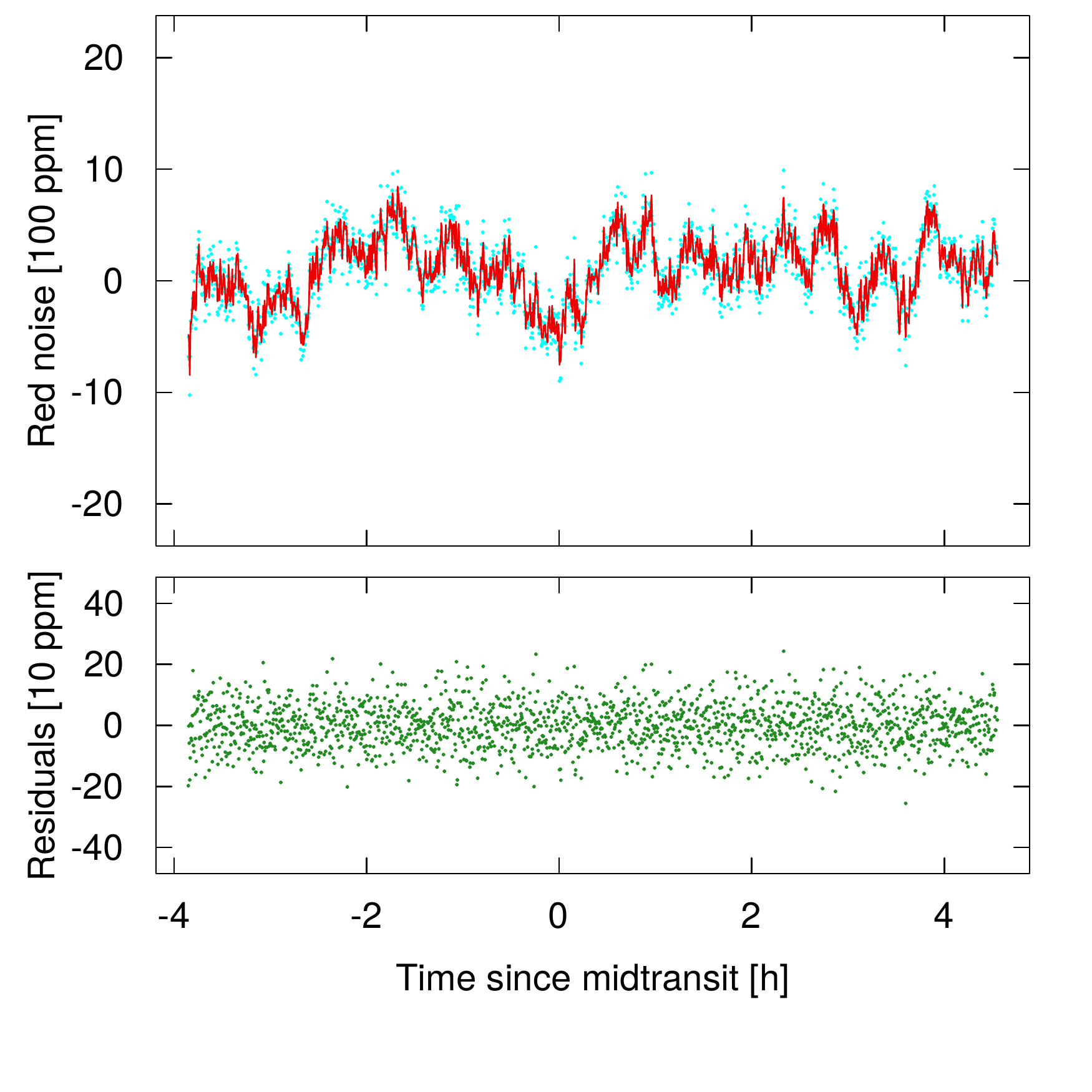}
    \caption{Same simulated light curve shown in Fig. \ref{fig:typic} (upper-left panel, in red), along with the fitted transit light curve via TLCM (upper-left panel, in blue) and the residuals (lower-left panel). Note: the TTV-like phenomena is not present in this case, as is visible on the two inset plots. The upper-right panel shows the residuals (or, ideally, the identified noise) with cyan, while the red curve denotes the fitted correlated noise. The residuals of the noise fitting are shown on the bottom-right panel. }
    \label{fig:red}
\end{figure*}

The second code, TLCM, utilises a generic algorithm to minimise the logarithmic likelihood, a simulated annealing refines the fit, and MCMC is used to estimate the error bars \citep{2020MNRAS.496.4442C}. In this fitting procedure, $t_C$, $p$, $a/R_S$, and $b$ were treated as free parameters, while in this case, $P$, $u_1$, and $u_2$ were set to their input values as well. As the limb darkening was known a priori, we chose to fix $u_{1,2}$ in both fitting codes, even though \cite{2013A&A...549A...9C, 2020AJ....159..123A} suggests that leaving them as free parameters would increase the precision of the planetary radii. However, TLCM allows for the handling of the correlated noise using the wavelet formulation routines of \cite{2009ApJ...704...51C}, which are built from the Daubechies fourth-order wavelet basis \cite{daub} . For further details, we refer to Paper I. These describe the red noise in terms of $\sigma_w$ (white component) and $\sigma_r$ (red component). As this method is more time-consuming, we solved only 1,500 light curves using this procedure. For a tentative comparison, Fig. \ref{fig:red} left column shows the same simulated light curve as Fig. \ref{fig:typic}. We note that the analysis of the nature of the correlated noise component is beyond the scope of this paper.

There are several other applications of wavelets in exoplanetary science. \cite{2010SPIE.7740E..19W} and \cite{Li:DVmodelFit2019} performed model fits to transit light curves in the wavelet domain in the context of the Data Validation of the Kepler Science Operations Control and the performance of the Kepler Science Data Processing Pipeline in relation to limb-darkened transit model-fitting and multiple-planet search algorithms. \cite{2002ApJ...575..493J} used a wavelet-based approach to characterise the correlated noise processes imposed onto transit light curves of the Kepler light curves \citep{2020TPSkdph} due to stellar variability.

\subsection{Analysis of the resulting datasets: Statistical moments}

The distributions of the datasets compiled in this way were analyzed primarily through their statistical moments. The third and fourth central moments for a univariate data, $X_1, X_2, \dots, X_N$, is given by:
\begin{equation}
    g_\mu = \frac{\sum_{i=1}^N \left(X_i - \overline{X} \right)^\mu/ N}{\sigma^\mu},
\end{equation}
where $\overline{X}$ is the mean  (the first moment), $\sigma$ is the standard deviation (the positive square root of the second central moment, the variance), and $\mu \in \{3,4\}$. 

In the case of $\mu=3$, $g_3$ is called the skewness of the distribution. As the skewness of a normal distribution is zero, negative skewness results from data that are skewed left (i.e. the left tail is long relative to the right tail), while a positive value for $g_3$ means data that are skewed right (i.e. the right tail is long relative to the left tail).

In case of $\mu=4$, $g_4$ is the kurtosis of the distribution. The kurtosis of a Gaussian distribution is $3$, meaning that kurtosis values larger than $3$ indicate data that is heavy-tailed, while a kurtosis lower than $3$ implies a light-tailed distribution. 

\section{Results} \label{sec:res}

To examine the distortion caused by the red noise, we calculated three sets of solutions with the different fitting algorithms and compared the best-fit values of $t_C$, $p$, $b^2$, and $\omega$ that were the result of different noise models or fitting algorithms. The noise characteristics (correlated and uncorrelated parts) were chosen to be identical. As we made use of a simulated transit light curve, we knew the true values of the fitted parameters and we were able to compare the actual differences to the error terms given by the fitting algorithms. We emphasise that prior knowledge of the transit parameters does not influence the fitting and error estimation processes. This is enforced by selecting wide fitting intervals (i.e. uniform priors) for each parameter.


First, we compared the results of the linear optimisation approach of \textit{lfit} for both noise models. Figure \ref{fig:abs_r_w} shows the distribution of the absolute errors of the fitted parameters, while Fig. \ref{fig:rel_r_w} 
compares the experienced errors (the difference of the input and fitted parameters) to the error estimate calculated by the algorithm itself. 
We then compared the correlated noise dataset from Figs. \ref{fig:abs_r_w} and \ref{fig:rel_r_w} to the wavelet-based light curve solutions of TLCM. The absolute and relative error distributions resulting from these two (generally different) fitting algorithms and their statistical momenta are shown in Figs. \ref{fig:TLCM_abs_r_w} and \ref{fig:TLCM_rel_r_w}.

There are two kinds of light curve solutions computed by TLCM: the statistical median within the region of acceptance (where all MCMC solutions are statistically indistinguishable from the data) and the single best-fit solution. We compared these solutions to test the robustness of our methods. Computing the cumulative distribution function for the two types of solution, the two-sample Kolmogorov-Smirnov test allows the comparison of the two distributions (see e.g. \citealt{ feigelson_babu_2012}). Exploring this statistic for the $t_C$, $p$, $b,$ and $a$ parameters, we may conclude that the resulting $p$-value of 1 means that the null hypothesis (that the two cumulative distribution functions are equal) is acceptable. The parameters presented in Figs. \ref{fig:TLCM_abs_r_w}. and \ref{fig:TLCM_rel_r_w}. are calculated from the median solution, however, this means that the selection of solutions should be of no consequence. 

\subsection{Time of midtransit}

The \verb|lfit| algorithm reproduces the experienced error of $t_C$ in the case of white noise only, but it has been found to be too optimistic when red noise arises. The distributions of the resulting $t_C$ values  can be seen in  Figs. \ref{fig:abs_r_w} -- \ref{fig:TLCM_rel_r_w} (upper-left panels).

The comparison between the two cases (white noise only and red noise only)  solved via \verb|lfit| yields the following results (Figs. \ref{fig:abs_r_w}, \ref{fig:rel_r_w}). The median of the distribution of the absolute error in case of the uncorrelated noise model is $0$ and in the red noise cases, it is $\sim 2.6$ seconds (within the estimated uncertainty of $6.2$ seconds and it is, therefore, insignificant). For both histograms, the presence of correlated noise increases the variances by roughly an order of magnitude (by a factor of $17.7$ and $23.7$). Visually, this appears as a broader histogram with long 'tails', where the parameters describing the fitted transit light curve have values much different than the original ones.  
For a Gaussian distribution, the variance of the relative error distribution is unity and, indeed, for the white noise case, this has been reproduced. However, in the red noise case, we may conclude that the uncertainties are $\sim 5.4$ times underestimated by \verb|lfit| on average. The negative values of the skewness of every distribution suggest that $t_C$ tends to be overestimated, meaning that the fitted light curves are shifted more often to the right of the true midtransit. 

Comparing the wavelet-based approach of TLCM to the white assumption of \textit{lfit} while using the same red noise model for the simulations (Figs. \ref{fig:TLCM_abs_r_w}, \ref{fig:TLCM_rel_r_w}), we find that the median of these absolute and relative error distributions is also $\sim 0$. While the variance of the absolute distribution is about $17.7$ times lower than for the case of the \textit{lfit} data, the proper error estimation of the wavelet-formulation results in a variance that is $\sim 9.6$ times lower where the relative errors are concerned. The skewness is also negative in these cases, while the kurtosis of this more complex approach is lower for both the absolute and relative errors ($4.7$ and $1.1$ times, respectively). Visually, this translates to a lack of elongated tails, and in the case of Fig. \ref{fig:TLCM_rel_r_w}, it exhibits a narrower distribution. 
We note that the small discrepancies between the values of the statistical momenta shown in Figs. \ref{fig:rel_r_w} and \ref{fig:TLCM_abs_r_w} are due to the differences in bin sizes.  

\subsection{Ratio of the planetary and stellar radii}

The top right panels of Fig. \ref{fig:abs_r_w} -- \ref{fig:TLCM_rel_r_w} show the absolute and relative errors for the fitted values of $p$.
In examining the differences between the effects of the white and correlated noises when only the white noise is accounted for (Fig. \ref{fig:abs_r_w} and \ref{fig:rel_r_w}), we find a similar behaviour to what was observed for $t_C$. All four distributions share a median value $\sim 0$ (the value of $10^{-5}$ is lower than the average estimated uncertainties of $20.1 \cdot 10^{-5}$). However, the reduced variances in the parameter determined by the \verb|lfit| algorithm is increased by a factor of $18.0$ and $27.3$ when the red noise appears (and we, of course, do not account for it). Combining this with the kurtosis values
, we find that fitting the light curves simulated using correlated noise results in more outliers. The negative skewness mean that, for both noise models, $p$ tends to be underestimated more often via \textit{lfit}.      

The benefits of the wavelet-based approach can clearly be seen from the data in Fig. \ref{fig:TLCM_abs_r_w} and \ref{fig:TLCM_rel_r_w}. The median resulting from the fitting done via TLCM ($-7.5\cdot 10^{-5}$) is negligible compared to the average uncertainties of $91.8 \cdot 10^{-5}$. The variances for the absolute and relative cases are lower by a factor of $1.5$ and $20.4$, respectively. The negative values for the skewness of both distributions suggest that the radii are also overestimated more often. The values of the kurtosis ($1.707$ and $1.625$), along with the higher values for the variance, suggest that the white assumption leads to a broader distribution consisting of more outliers. The comparison of the distribution from the untreated correlated noise to a Gaussian distribution suggest that the estimated uncertainties are underestimated by a factor of $5.4$ as well.

\subsection{Impact parameter}

The absolute and relative errors of $b^2$ are shown in the bottom left panels of Fig. \ref{fig:abs_r_w} -- \ref{fig:TLCM_rel_r_w}. Here, we witness a very similar behaviour to that of the already discussed parameters.

The difference between the untreated correlated noise and the purely white noise models are similar to the cases of $t_C$ and $p$ (Fig. \ref{fig:abs_r_w} and \ref{fig:rel_r_w}). The median of all four distributions is $0$, while the variance is lower by a factor of $19.4$ and $24.0$ for the light red noise cases. The negative skewness values 
for the distribution of the absolute and relative errors for the two distributions suggest that with both noise types, the impact parameter tends to be overestimated. In both distribution types, the kurtosis is lower for the correlated noise cases ($10.760$ and $2.030$ compared to $17.584$ and $2.751$). Visually, we see broader histograms with tails for the red noise when comparing the two noise models that are caused by outliers, where the fit did not converge to the true value of $b$.

Examining the cases of the treated correlated noise (Figs. \ref{fig:TLCM_abs_r_w} and \ref{fig:TLCM_rel_r_w}), we see that the median of these two distributions is $\sim 0$, however, this is the only case out of the four studied parameters where the wavelet-based handling of the red noise yields absolute errors that have a broader distribution (with a variance that is $1.3$ times higher). The handling of the red noise by TLCM leads to better estimated uncertainties, meaning that the relative errors have a narrower distribution, where the variance is $35.1$ times higher for the linear pattern detection cases. As the skewness of the distribution constructed from the wavelet-based solutions is also negative, 
we may conclude that the impact parameter has a tendency to be overestimated. The kurtosis values of the wavelet-based result of $1.808$ and $1.688$ also suggest that there are more outliers with the white assumption. In comparison to a Gaussian distribution, we find that for the latter case, the uncertainties are underestimated by about $4.9$ times. 

\subsection[Omega parameter]{$\omega$ parameter}

The distributions of the absolute and relative errors of the $\omega$ parameter (proportional to the transit duration) for the two noise models and two different fitting approaches are shown on the lower right panels of Figs. \ref{fig:abs_r_w} -- \ref{fig:TLCM_rel_r_w}.

The comparison between the light curve solutions from the two noise models when solving the resulting light curves with a software based on white assumption yields similar results to the ones seen with the previous parameters, as in the $\sim 0$ medians are accompanied by variances that are $11.3$ and $24.4$ times larger for the correlated cases (the value of $6\cdot 10^{-3}$ d$^{-1}$ is negligible compared to the average uncertainties of $0.03$ d$^{-1}$). The negative skewness 
suggest that the $\omega$ parameter is also underestimated more often. The lower values for the kurtosis of red noise error distributions 
combined with the larger variances, suggest a broader distribution with more outliers.

Comparing the results from the wavelet-based approach, we also see similar features to those seen before (Fig. \ref{fig:TLCM_abs_r_w} and \ref{fig:TLCM_rel_r_w}). The median is also $\sim 0$ and the variances are $2.5$ and $287.7$ times larger for the cases where the noise is not modelled. This latter value suggests that the uncertainties from the wavelet-based approach are overestimated; this is most likely down to error propagation and, as a result,  the comparison of the relative errors should be treated carefully. The negative skewnesses of $-0.427$ and $-0.337$ also mean that the $\omega$ parameter tends to be underestimated. The narrower, tailless distributions of the TLCM results correspond to kurtosis values of $1.333$ and $0.926$. When comparing the red noise cases with linear pattern detection to a Gaussian distribution, we find that the uncertainty ranges are $\sim 5.1$ times underestimated.

\subsection{Possible application: Transit-timing variations}
\begin{figure*}[!ht]
    \centering
     \includegraphics[width=0.3\textwidth]{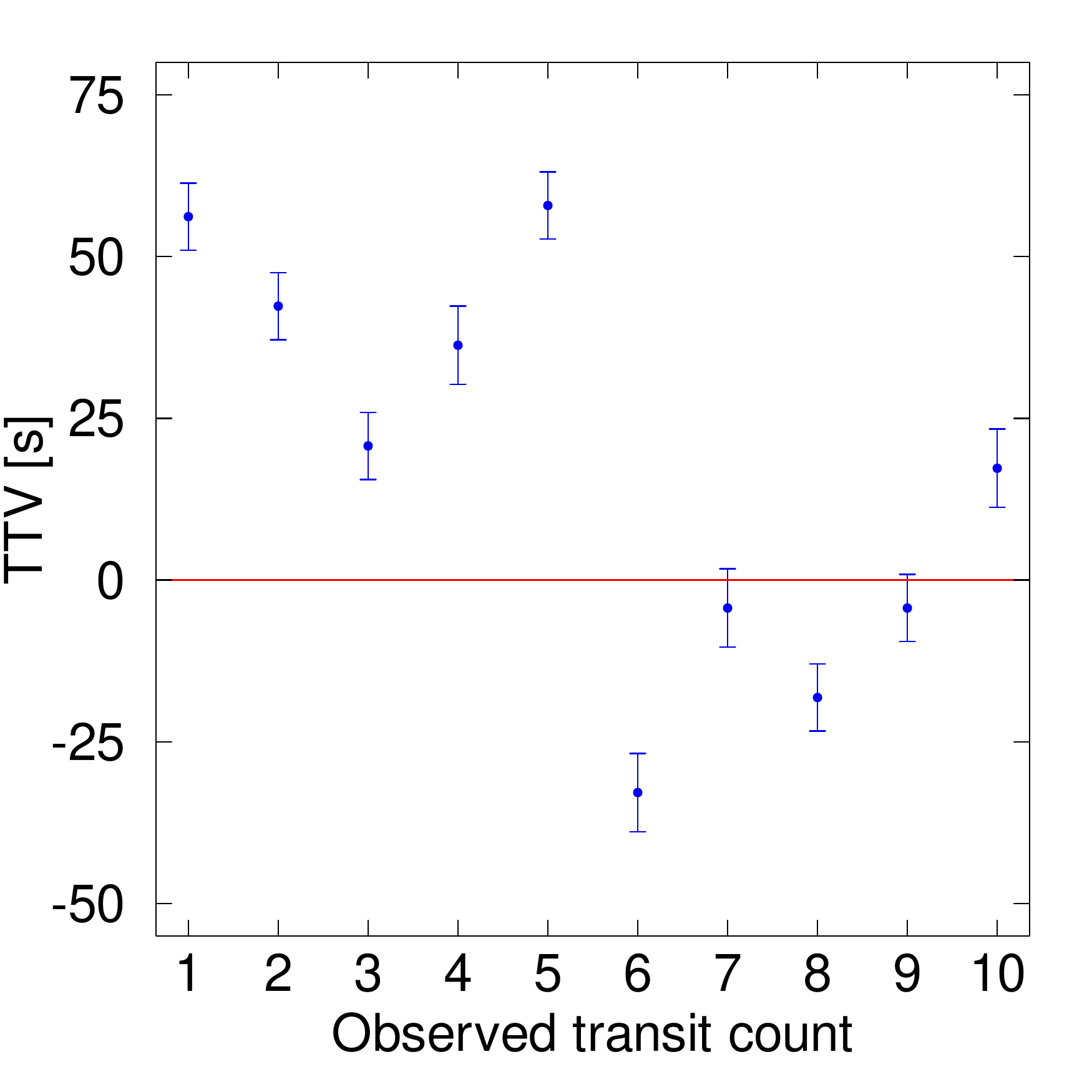}
    \includegraphics[width=0.3\textwidth]{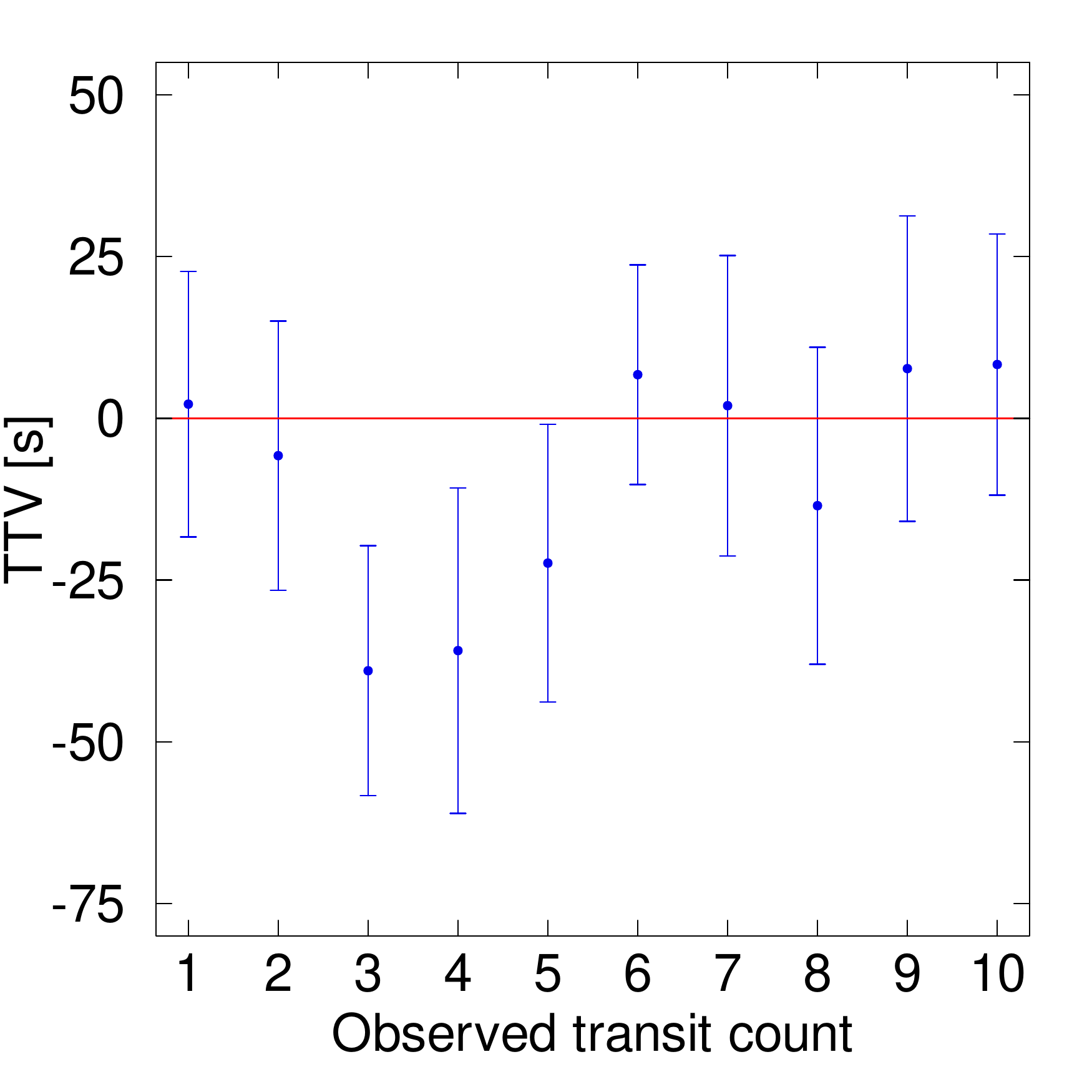}
    \includegraphics[width=0.3\textwidth]{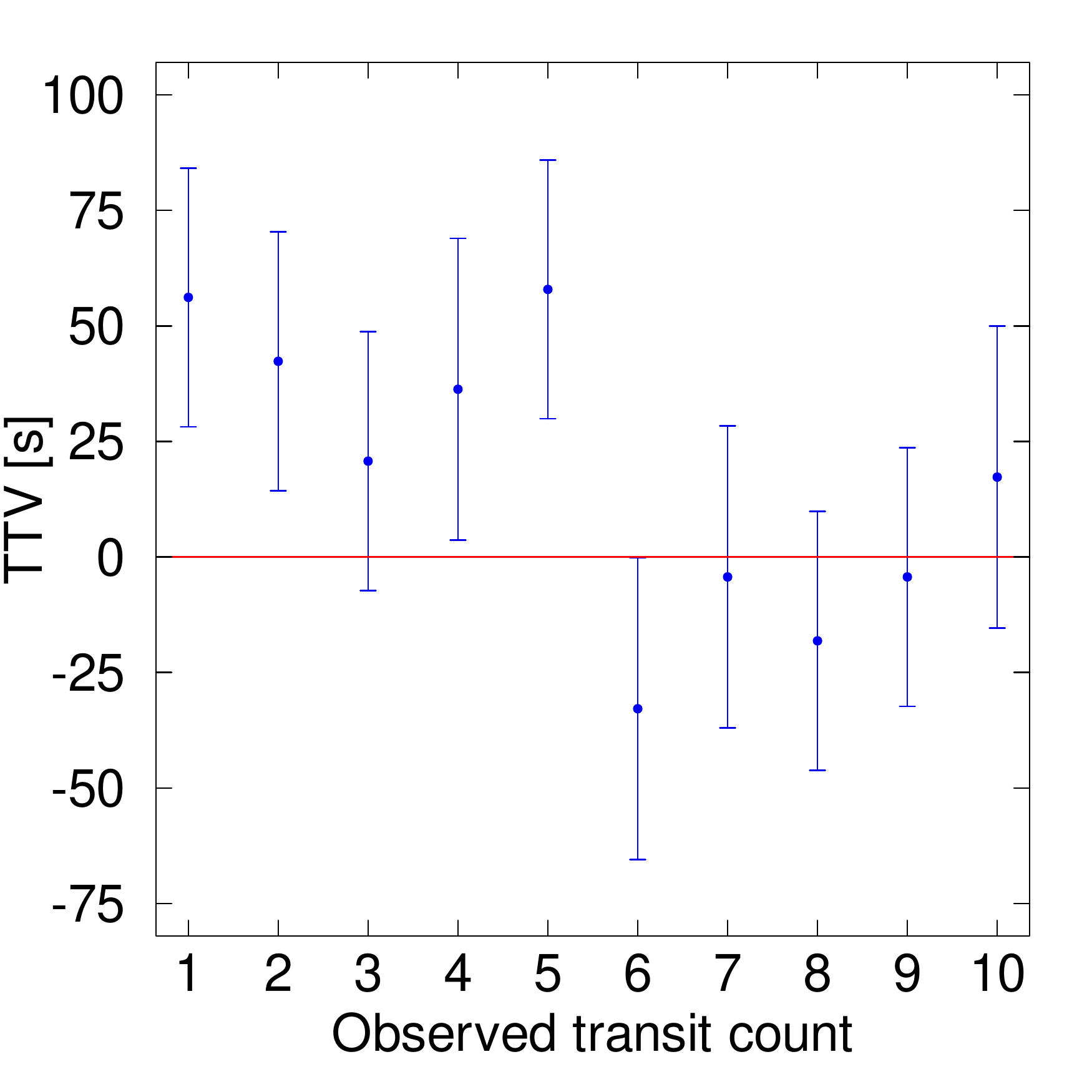}
    \caption{Simulated TTV detection that results purely from the red noise as there is no input TTV into to the signal. Left panel: Fitted $t_C$ values and the statistically estimated error bars from MCMC without handling the noise, resulting in a significant TTV signal. Middle panel: $t_C$ values after fitting the correlated noise through wavlet formulation, with no significant TTV. Right panel: Same as the left panel, but with error bars increased $5.4$ times, and with  no significant TTV.}
    \label{fig:ttv}
\end{figure*}

As we  discussed above, fitting a transit light curve with correlated noise  that is not handled properly displays two main flaws: (i) the fit does not necessarily converge to the true value of the parameters that describe the light curve and (ii) the error bars for said parameters tend to be severely underestimated. This could, for example, manifest itself in TTV-like phenomena (see Figs. \ref{fig:typic} and \ref{fig:red}). 

Taking a randomly selected segment from our fitted $t_C$ database and plotting it against a simulated multiple transit observation number (Fig. \ref{fig:ttv}, left panel), we get an apparent transit-timing variation. This effect is purely the result of the red noise, as we have no TTV in the input data. As discussed above, the statistical error estimate within the white noise assumption is too optimistic. When handling the noise with the wavelet approach, however, no such significant effect may be found (Fig. \ref{fig:ttv}, middle panel), meaning that the light curve solution is consistent. Scaling the error bars of the estimated time of midtransit by $5.4,$ as calculated from the bootstrap-like error estimation, the significance of the previously shown false TTV disappears (Fig. \ref{fig:ttv}, right panel). We note that, in reality, this number should be derived for each transit event individually. We interpret the consistency of the transit midtimes as a spectacular proof of the wavelet approach.

\section{Summary} \label{sec:sum}

In this paper, we present the benefits of the wavelet formulation as described in \cite{2009ApJ...704...51C} and implemented by the code TLCM as a means of analysing the transit light curves in the presence of time-correlated noise. We created a cloned noise model through an ARIMA process and used its different realisations for a bootstrapping-like approach in order to get a statistically significant amount of data of the parameters that characterise the exoplanet via the Mandel-Agol model \citep{2002ApJ...580L.171M}. We also fit multiple light curves with FITSH/\textit{lfit,} which does not account for the red noise.

Modelling the red noise allows for the proper estimation of the uncertainty ranges of each parameter. When comparing the histograms of the relative errors from the two different fitting procedures (Fig. \ref{fig:TLCM_rel_r_w}), we see a much narrower distribution in the TLCM-case, the variances of which are close to unity; in turn, this means that the error bars are correctly estimated. This suggests that modelling the correlated noise in this way provides reliable and consistent results for the light-curve solutions.

In order to answer the underlying question about the amount of distortion caused by the correlated noise, we compared the results to a white-noise model. We have found that without the proper treatment of the red noise, it biases the transit parameters significantly. This is manifested in two ways: (i) the distribution of the parameters is much broader in the correlated case and it has significant tails, meaning there is a heavy under- or overestimation of said parameters and (ii) the estimated error bars display an approximately five-fold underestimation. 

We have found, through extensive testing, that the simultaneous fitting of the transit itself and the red noise through a wavelet formulation can handle both these issues reasonably well. Our analysis may also be regarded as a way to estimate the uncertainties in the parameters describing the transit, without handling the noise within a given fitting software. We have also found that the effects of the correlated noise may mimic TTVs, making the treatment of red noise especially useful in this special case.

\begin{acknowledgements}
     This work was supported by 
     a PRODEX Experiment Agreement No. 4000137122 between the ELTE E\"otv\"os Lor\'and University and the European Space Agency (ESA-D/SCI-LE-2021-0025). Support of the Lend\"ulet LP2018-7/2021 grant of the Hungarian Academy of Science is acknowledged.

     SzCs thanks DFG Research Unit 2440: ’Matter Under Planetary Interior Conditions: High Pressure, Planetary, and Plasma Physics’ for support. He also acknowledges support by DFG grants RA 714/14-1 within the DFG Schwerpunkt SPP 1992: ’Exploring the Diversity of Extrasolar Planets’.
     
     Project no. C1746651 has been implemented with the support provided by the Ministry of Culture and Innovation of Hungary from the National Research, Development and Innovation Fund, financed under the NVKDP-2021 funding scheme.
     
    The authors gratefully acknowledge the European Space Agency and the PLATO Mission Consortium, whose outstanding efforts have made these results possible.

\end{acknowledgements}

\bibliographystyle{aa}
\bibliography{refs}

\begin{appendix}
\section{Histograms of the distribution of the examined parameters}

\noindent\begin{minipage}{\textwidth}

    \includegraphics[width=0.45\textwidth]{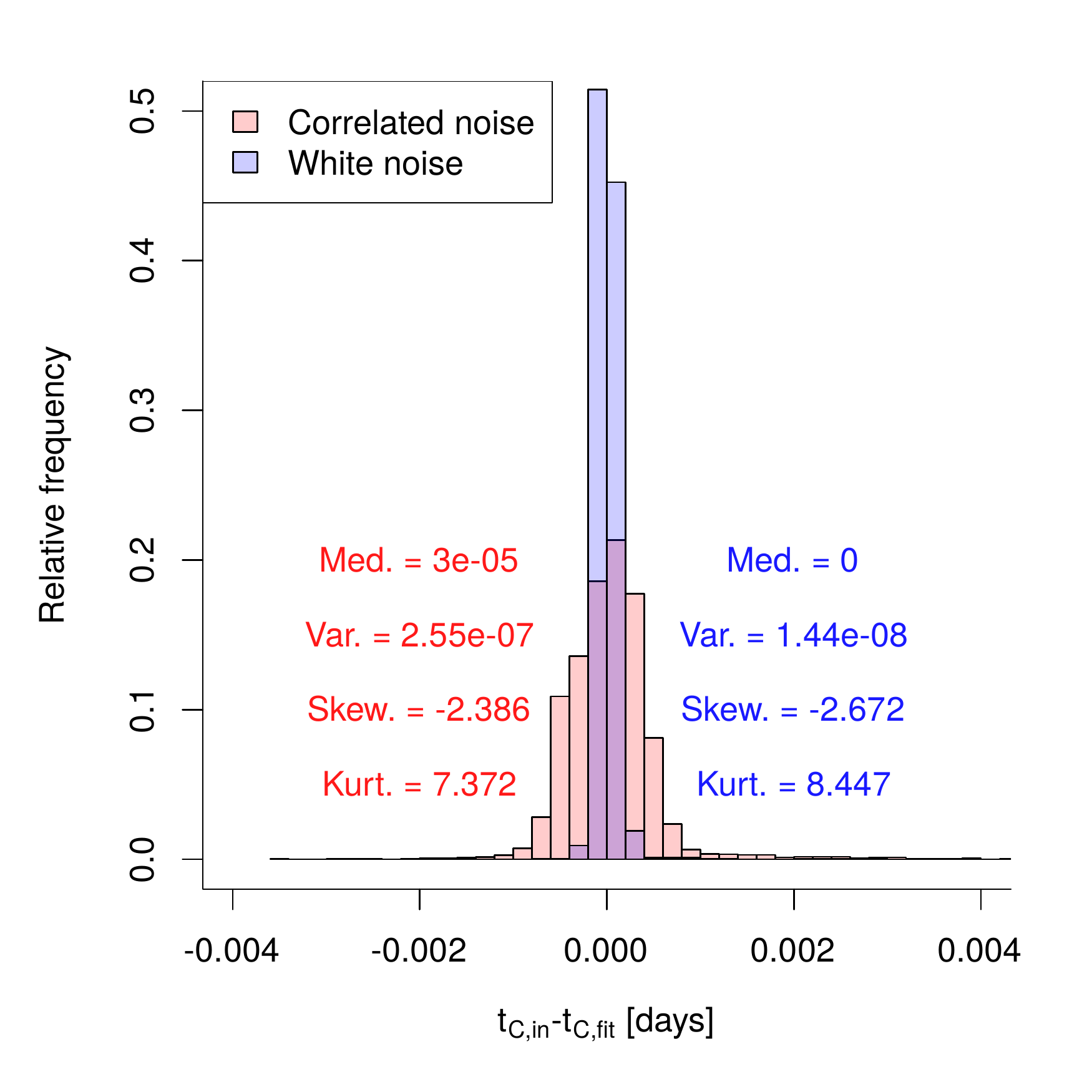}
    \includegraphics[width=0.45\textwidth]{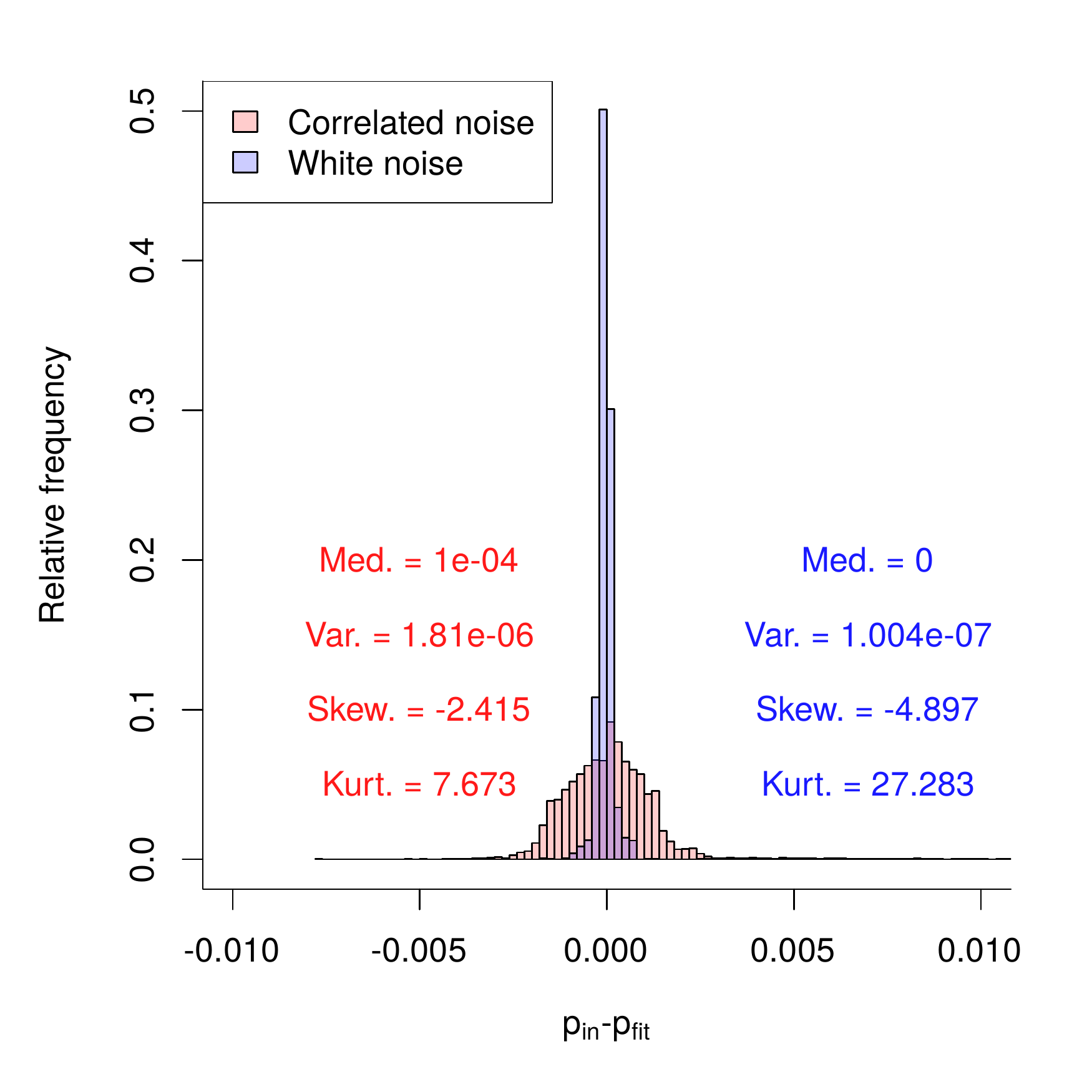}\\
    \includegraphics[width=0.45\textwidth]{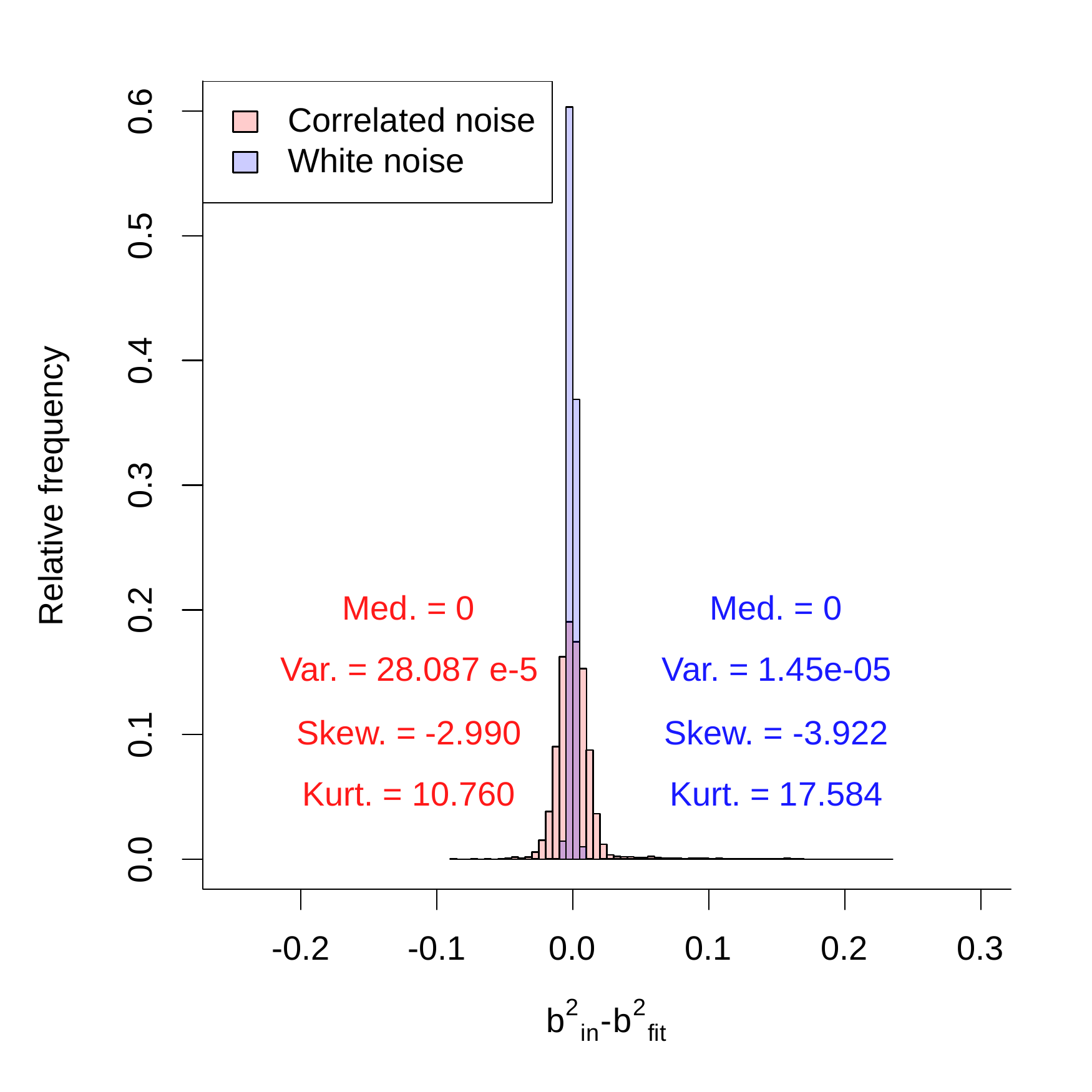}
    \includegraphics[width=0.45\textwidth]{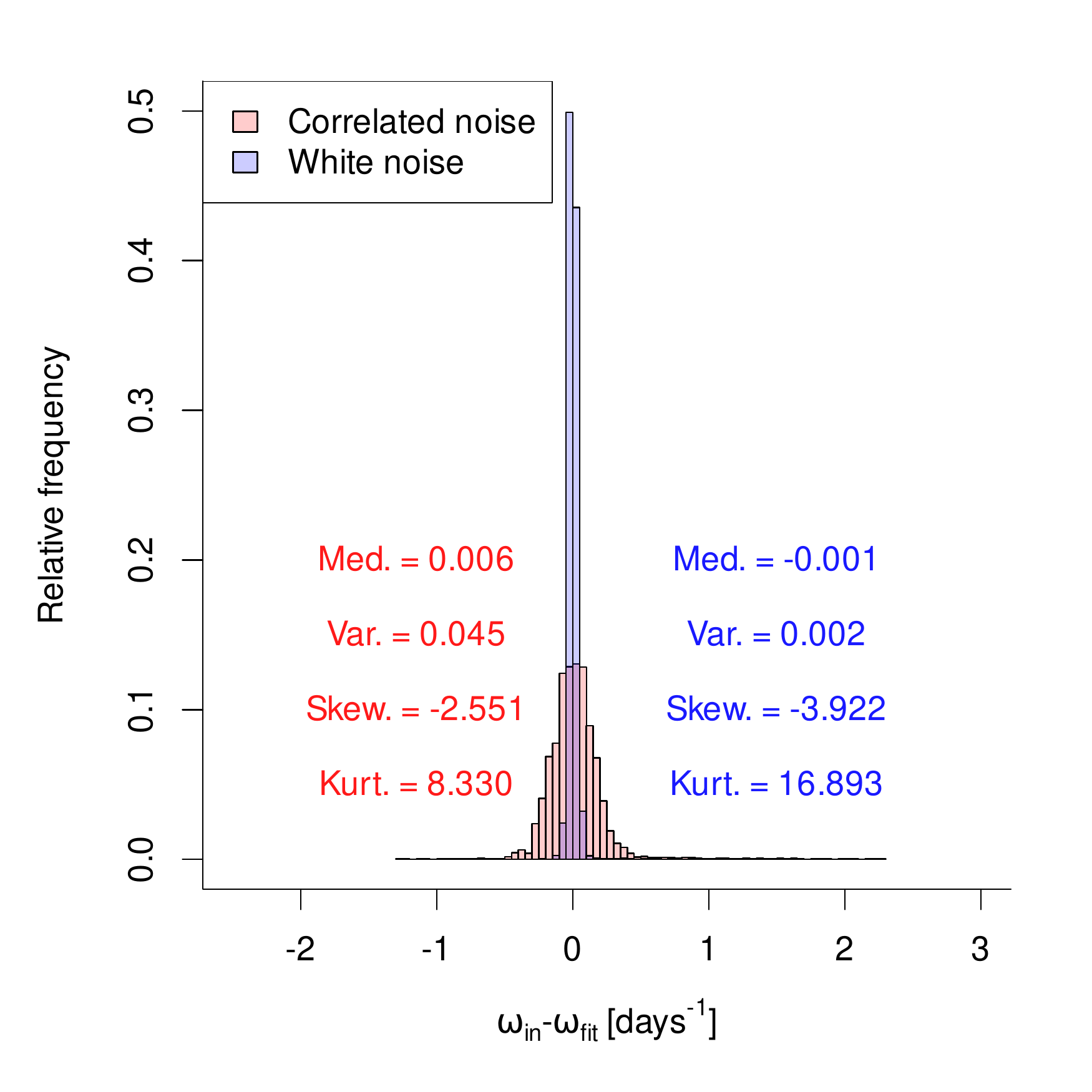}
    \captionof{figure}{Distribution of the differences between the input parameters and the fitted parameters for the red- and white-noise models, shown in red and blue, respectively. The median, variance, skewness, and kurtosis of the distributions is also shown.}
        \label{fig:abs_r_w}

\end{minipage}

\clearpage
\noindent\begin{minipage}{\textwidth}
    \centering
    \includegraphics[width=0.45\textwidth]{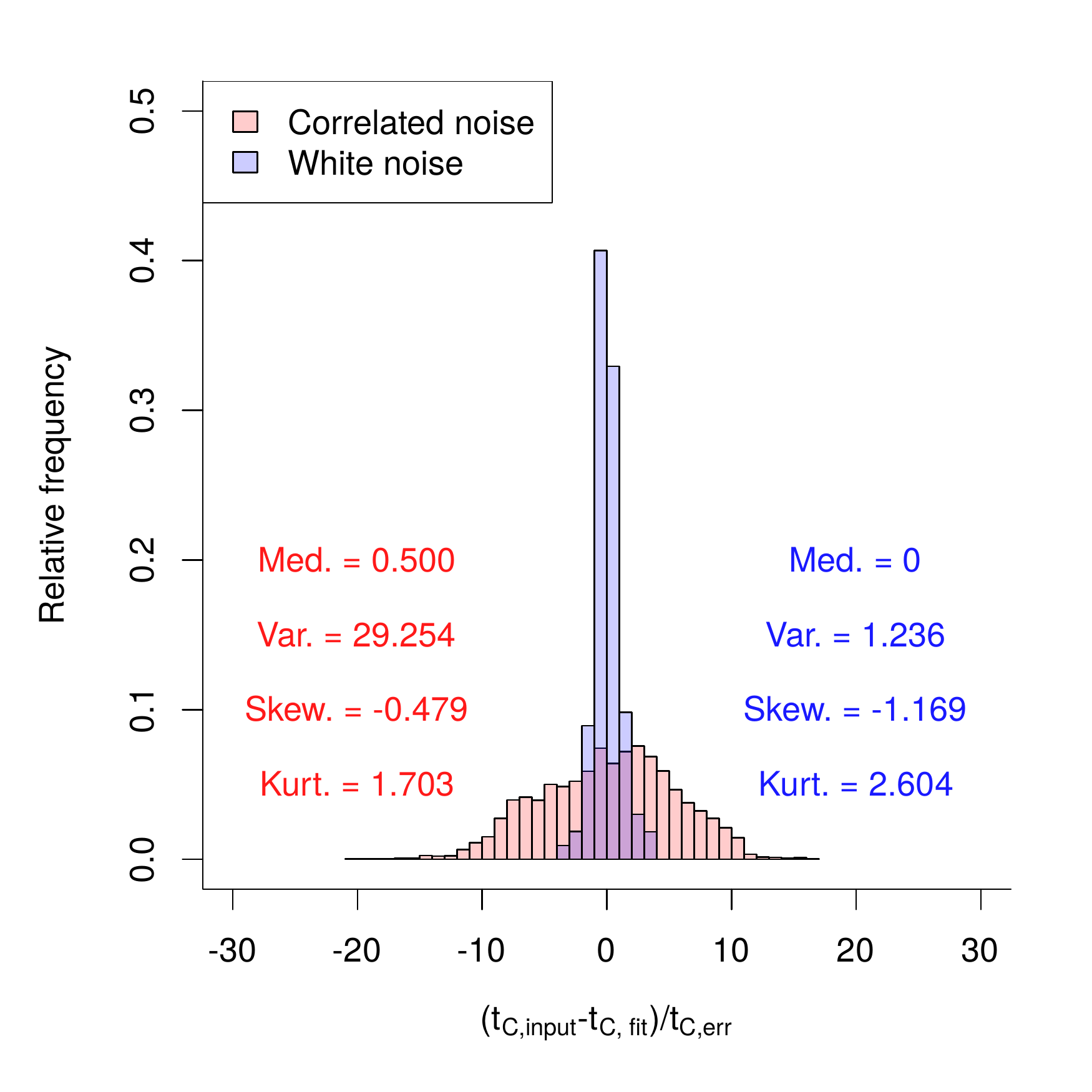}
    \includegraphics[width=0.45\textwidth]{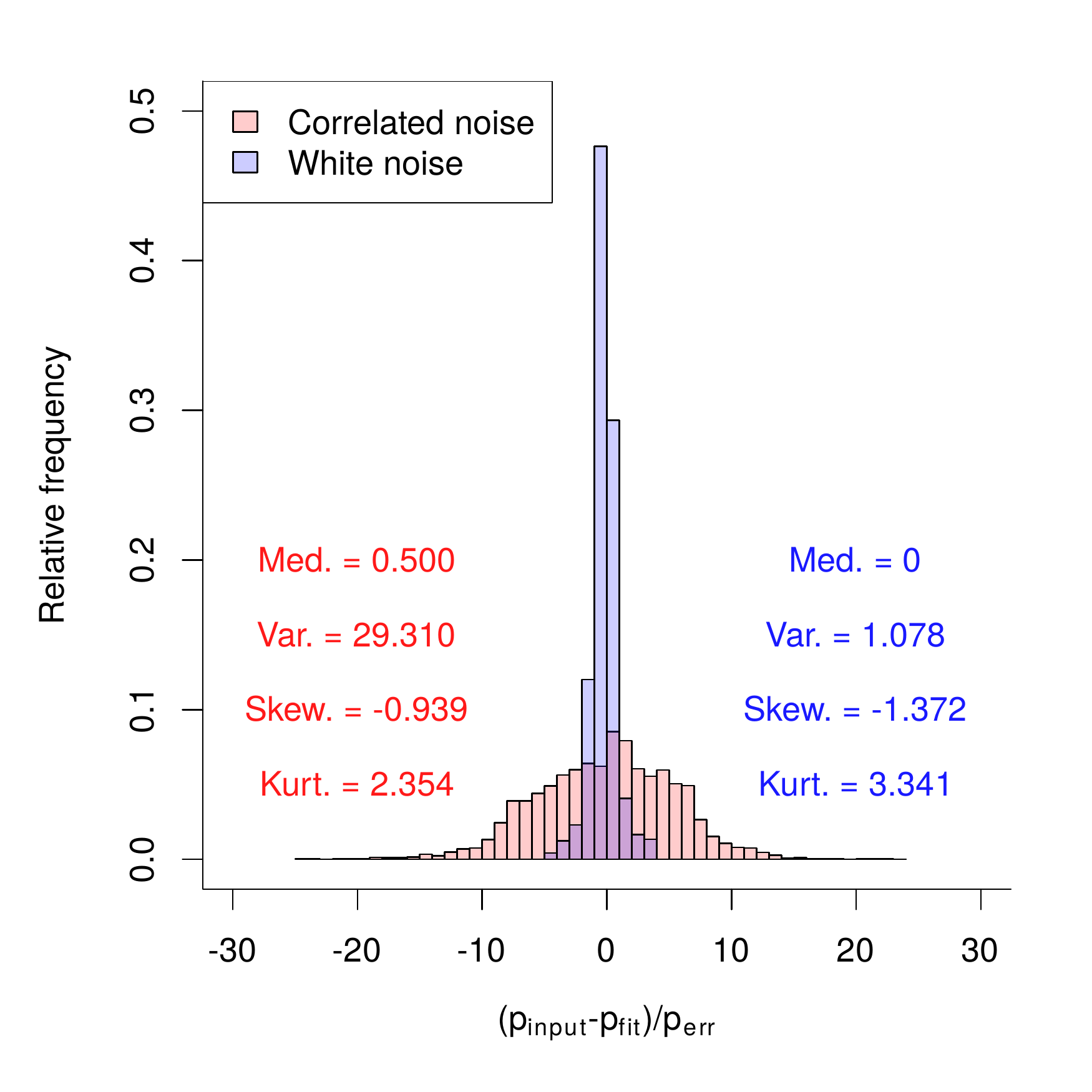}
    \includegraphics[width=0.45\textwidth]{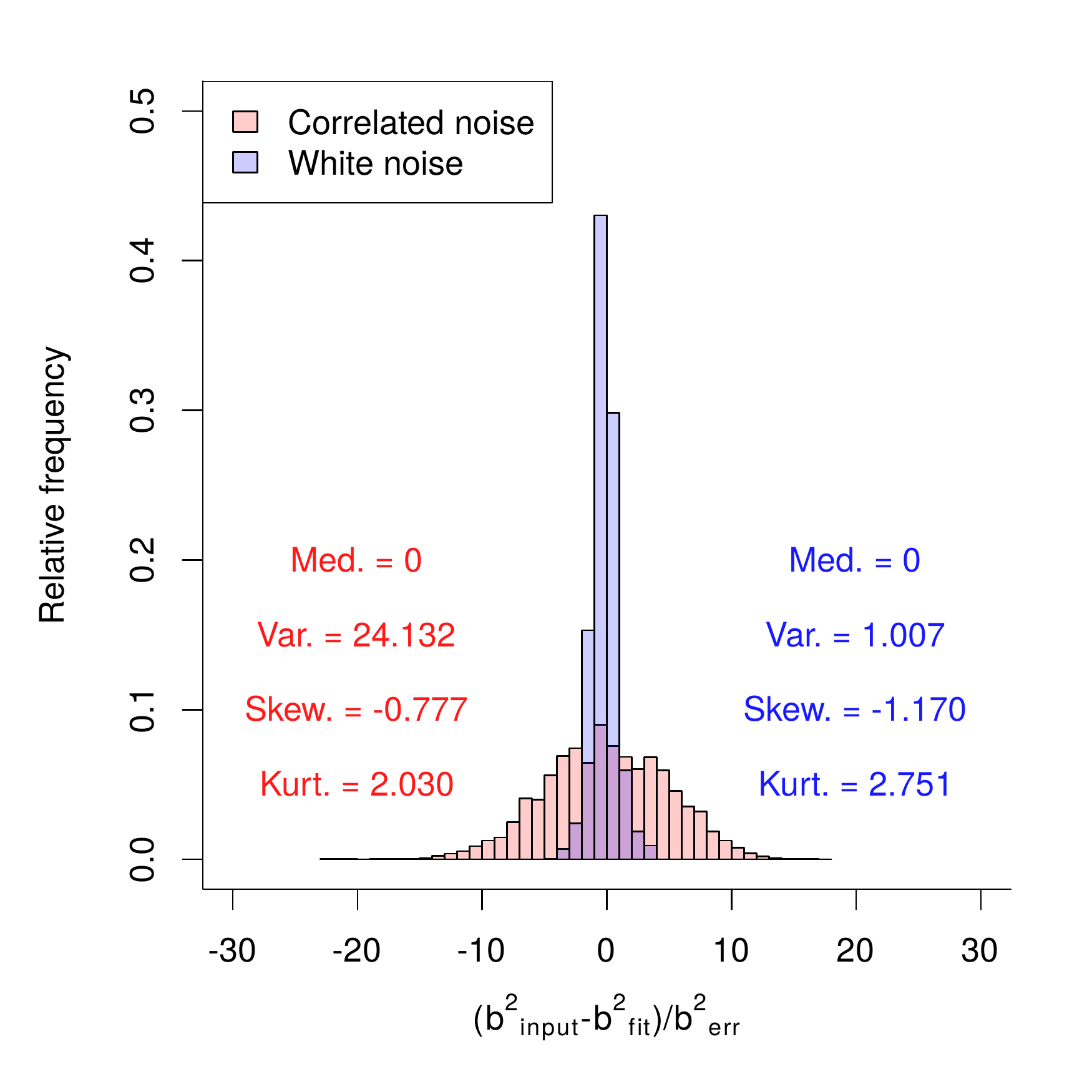}
    \includegraphics[width=0.45\textwidth]{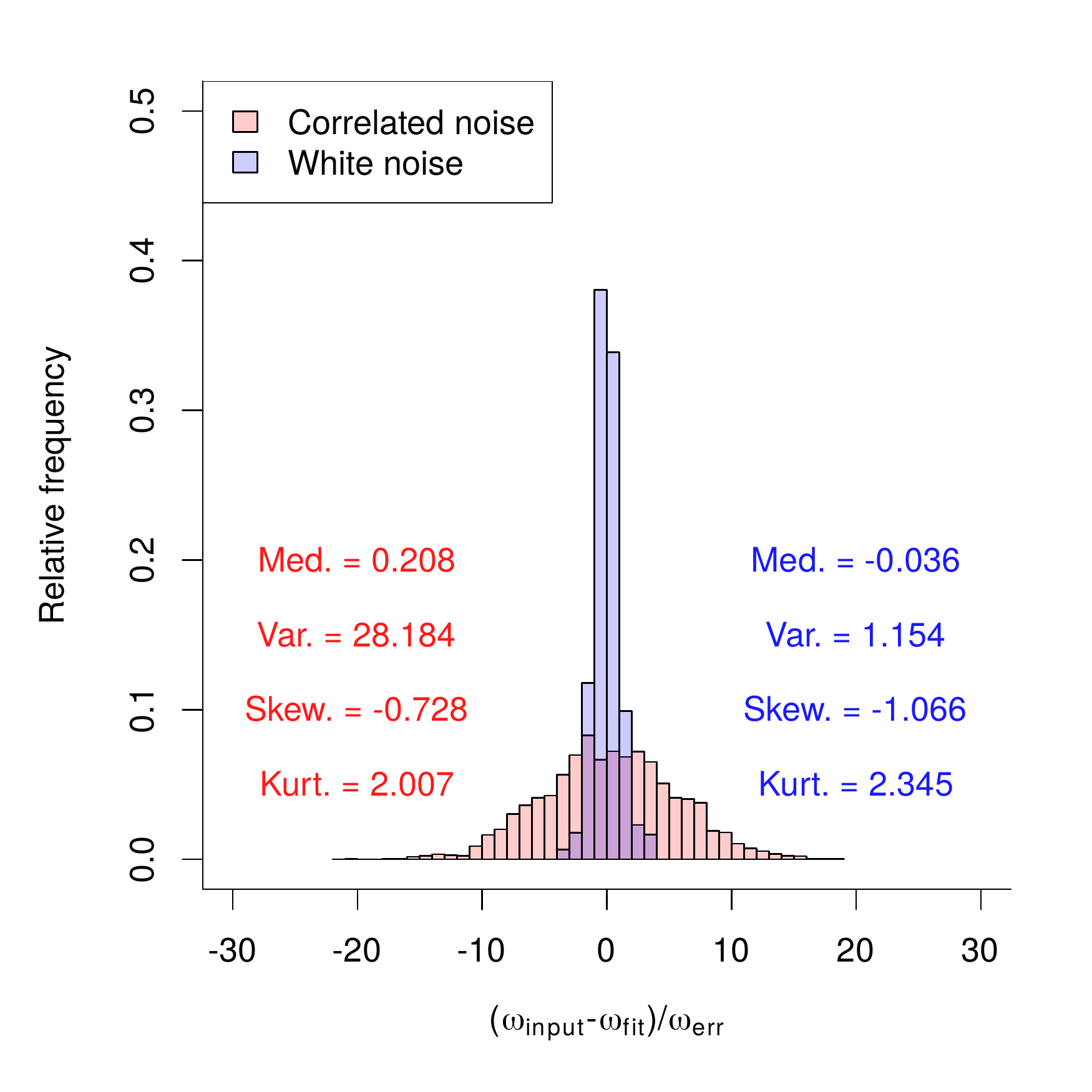}
    \captionof{figure}{Distribution of the differences between the input parameters and the fitted parameters, scaled with the estimated uncertainties, for the red- and white-noise models, shown in red and blue, respectively. The median, variance, skewness, and kurtosis of the distributions is also shown.}
        \label{fig:rel_r_w}
 \end{minipage}

\clearpage
\noindent\begin{minipage}{\textwidth}
    \centering
    \includegraphics[width=0.45\textwidth]{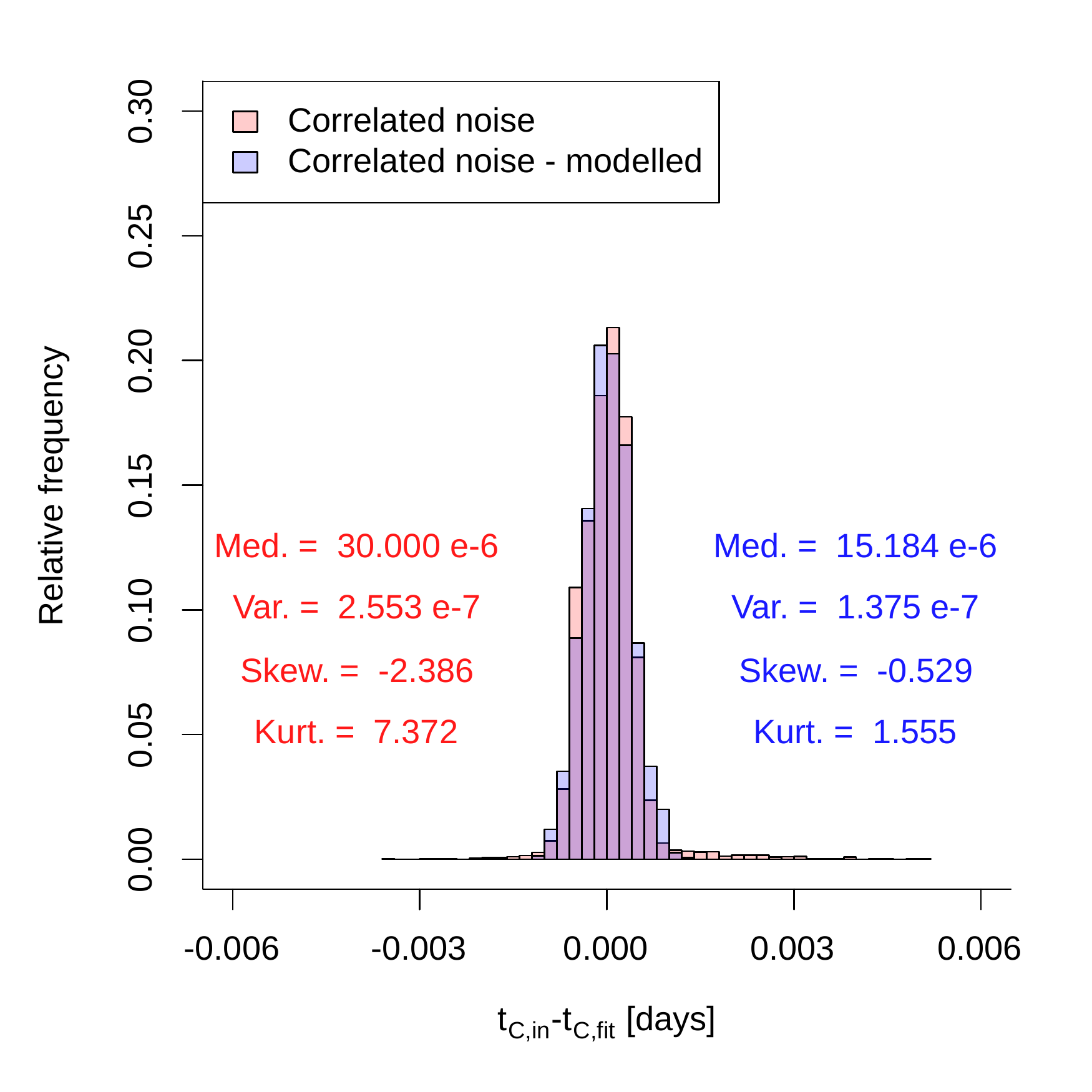}
    \includegraphics[width=0.45\textwidth]{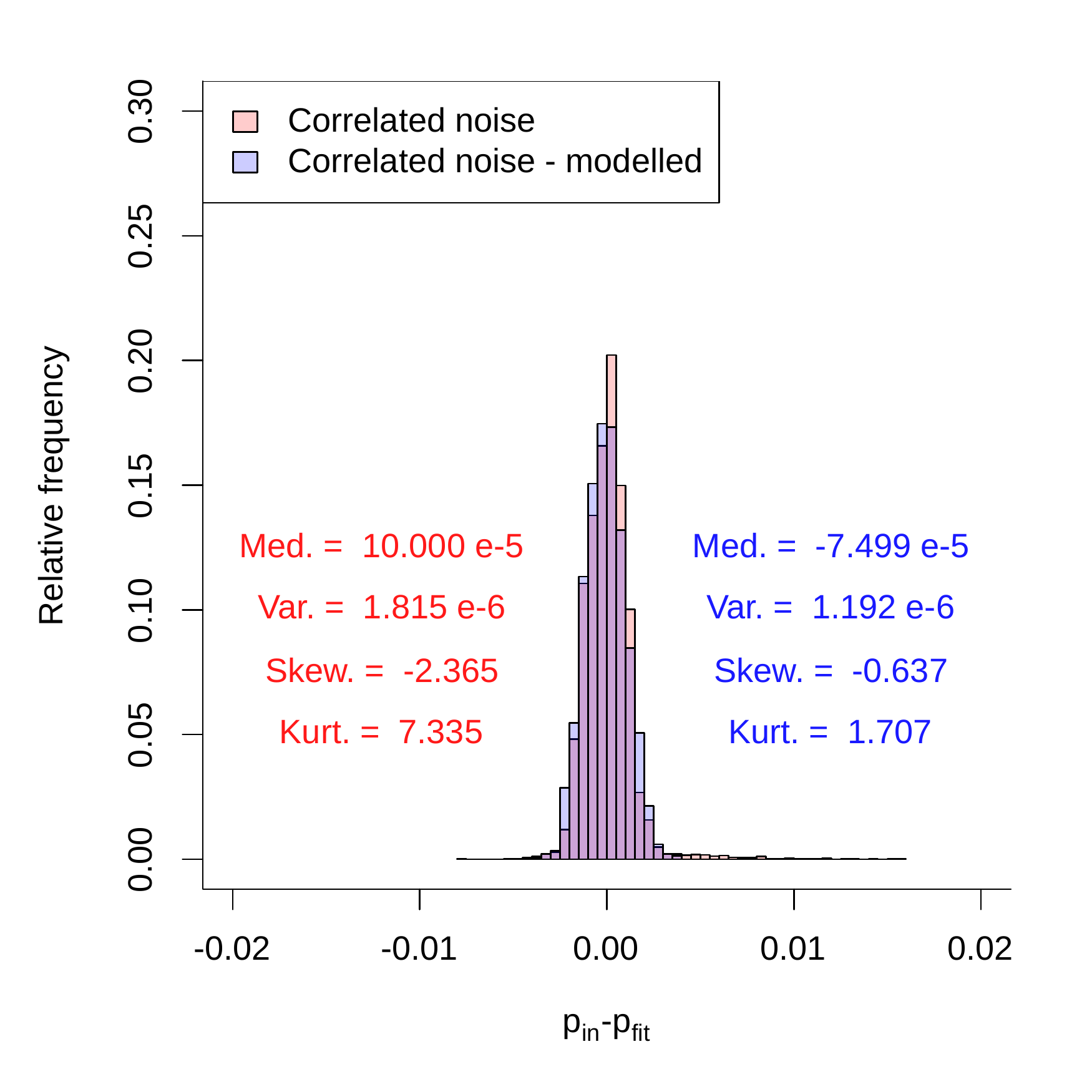}
    \includegraphics[width=0.45\textwidth]{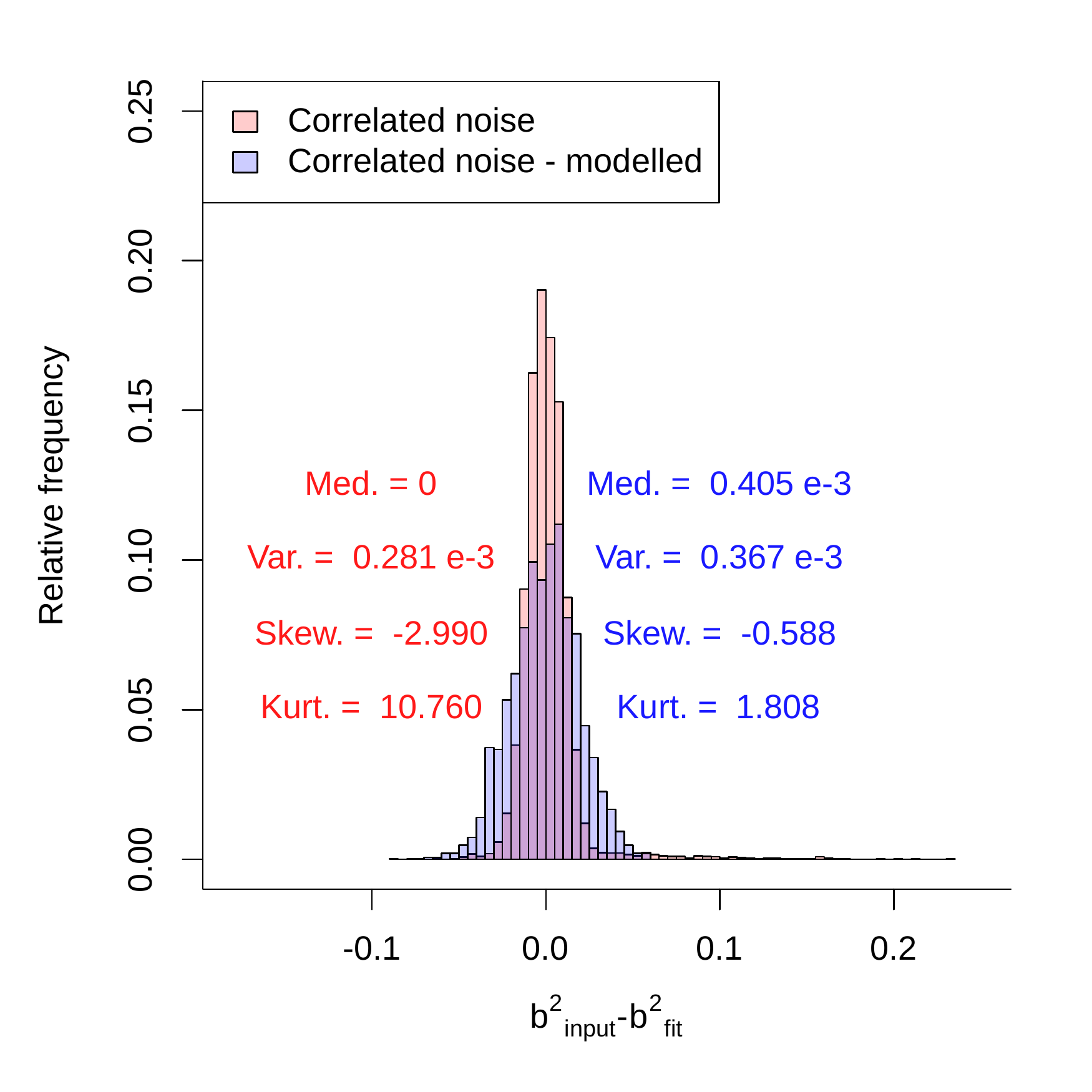}
    \includegraphics[width=0.45\textwidth]{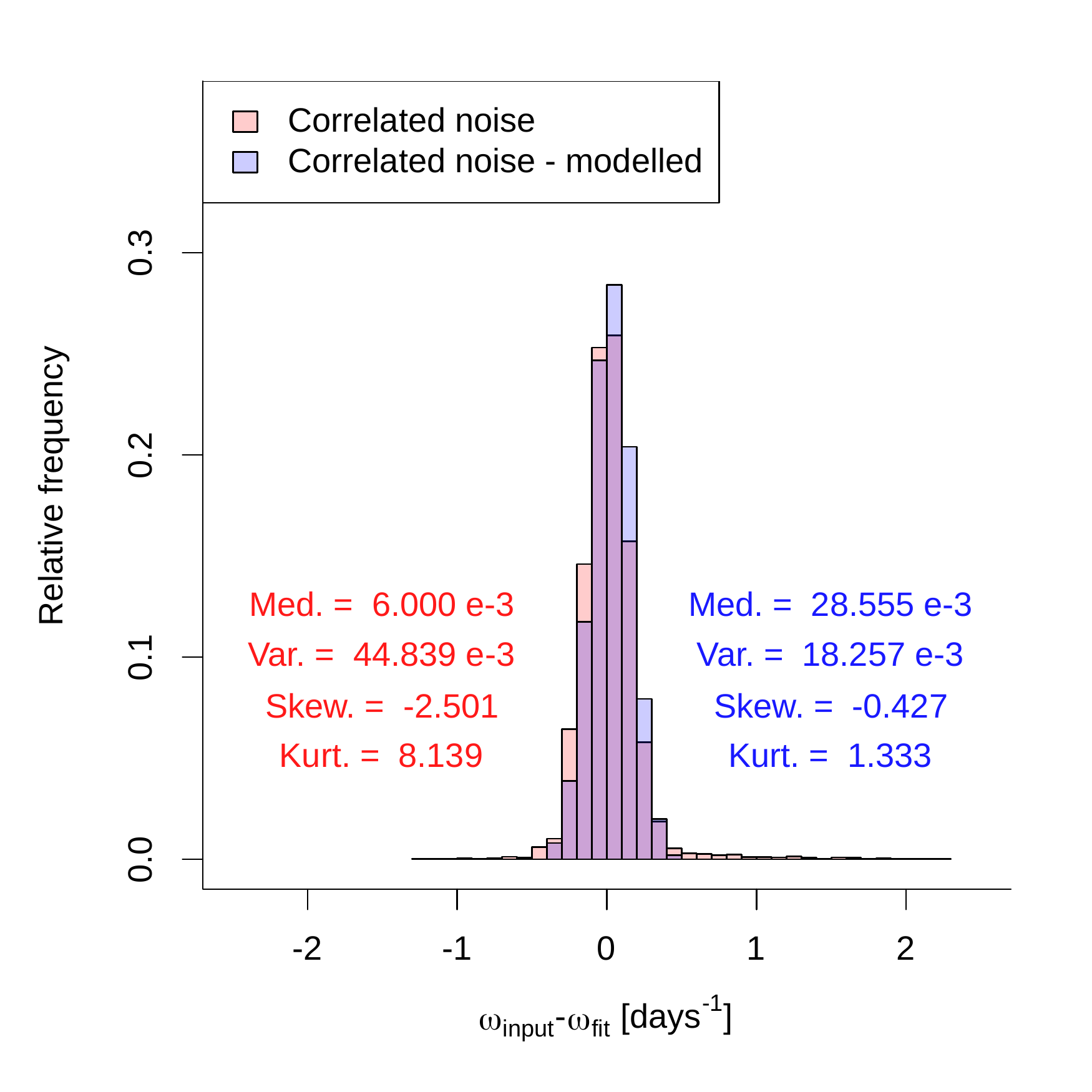}
    \captionof{figure}{Distribution of the differences between the input parameters and the fitted parameters for the same correlated noise model, without the noise fitting (red, i.e. same as in Fig. \ref{fig:abs_r_w}) and with the red noise handled through wavelet transformation (blue). The median, variance, skewness, and kurtosis of the distributions is also shown.}
    \label{fig:TLCM_abs_r_w}
 \end{minipage}

\clearpage
\noindent\begin{minipage}{\textwidth}
    \centering
    \includegraphics[width=0.45\textwidth]{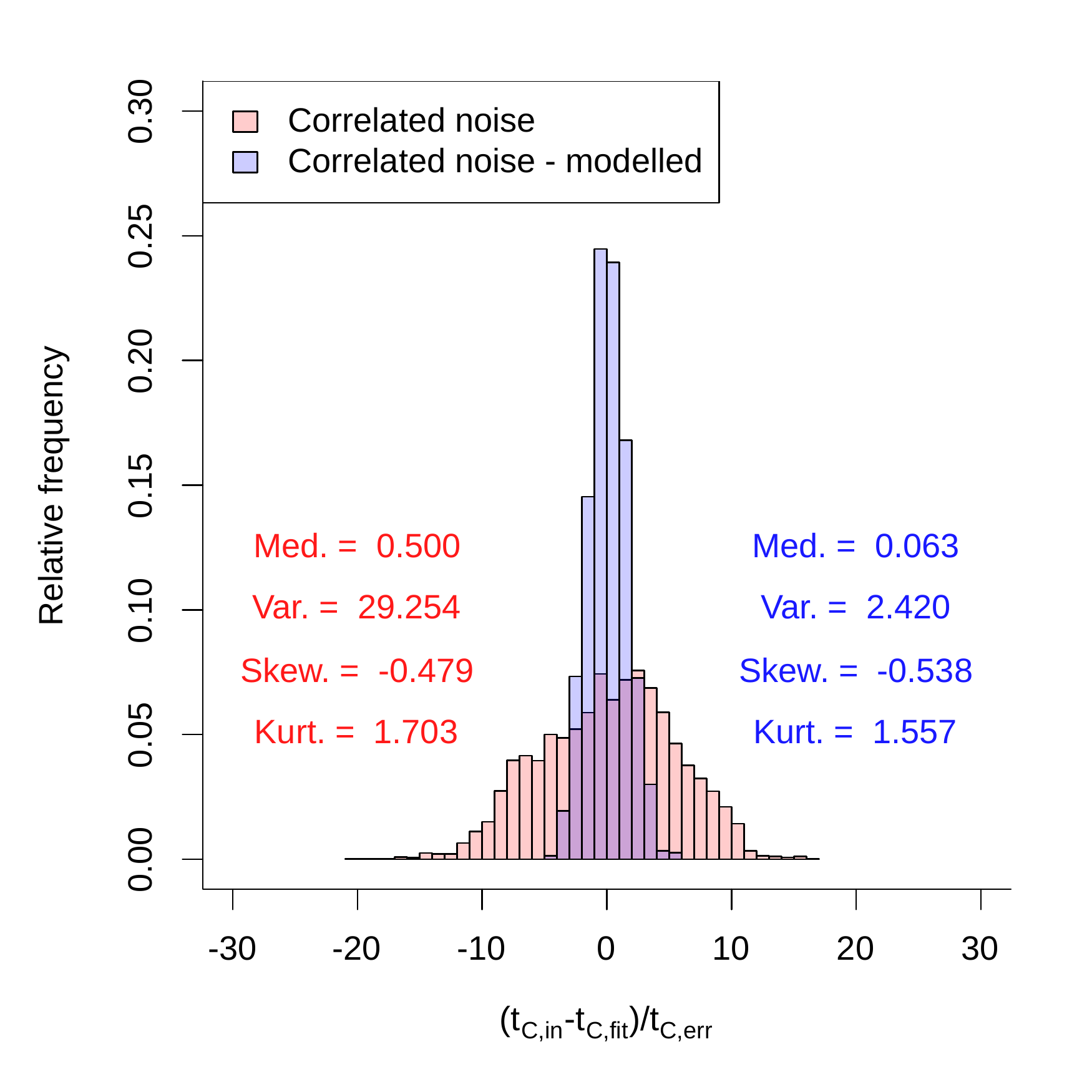}
    \includegraphics[width=0.45\textwidth]{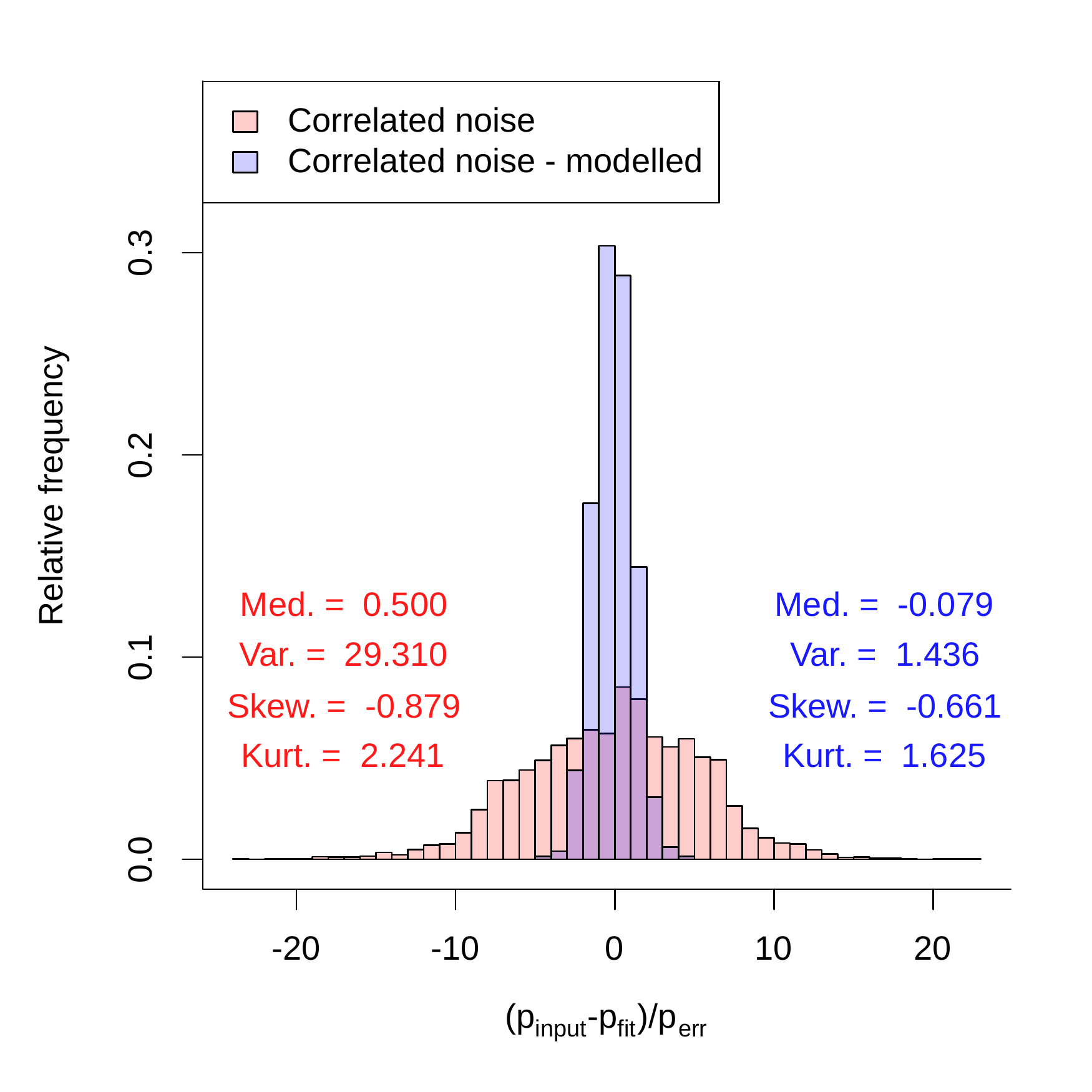}
    \includegraphics[width=0.45\textwidth]{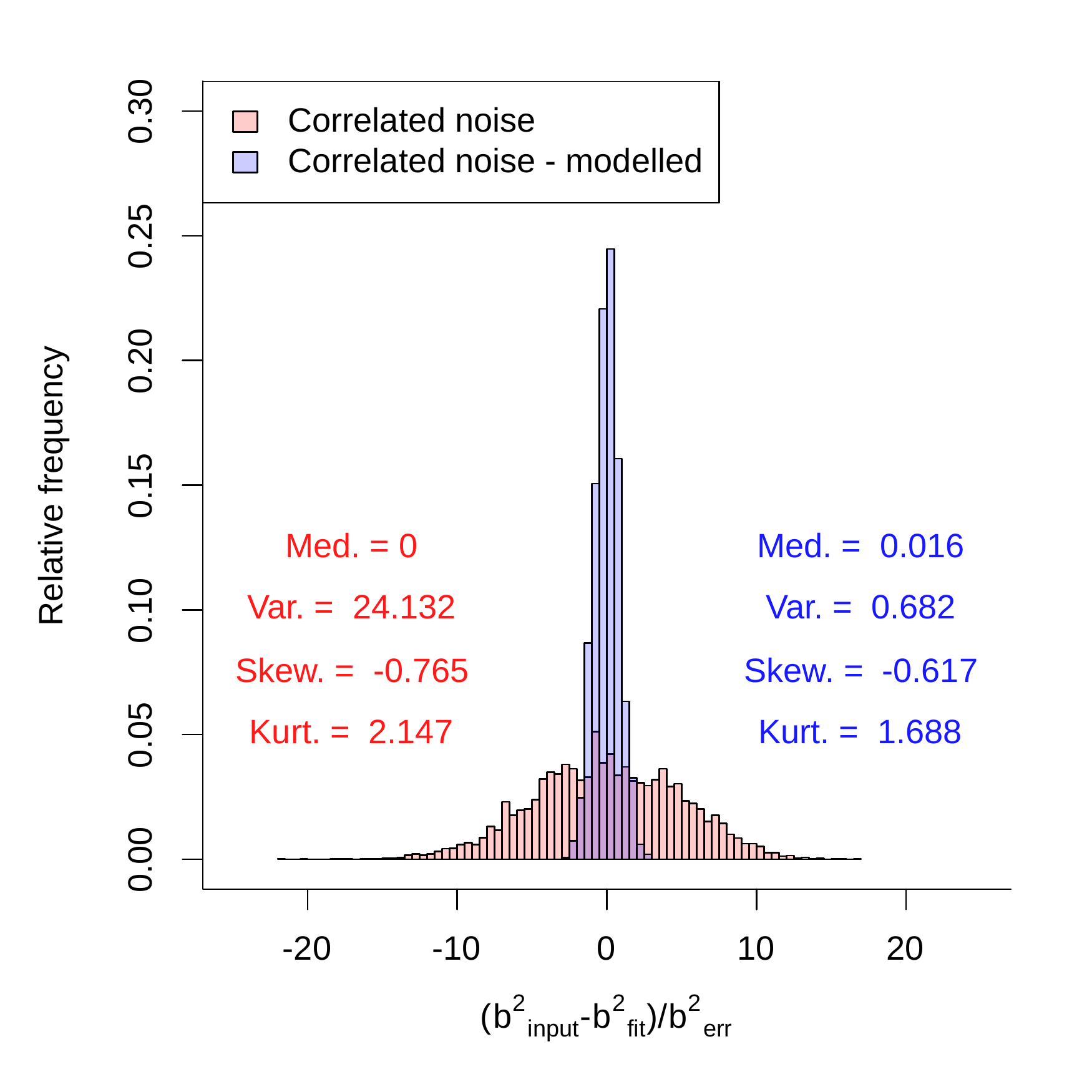}
    \includegraphics[width=0.45\textwidth]{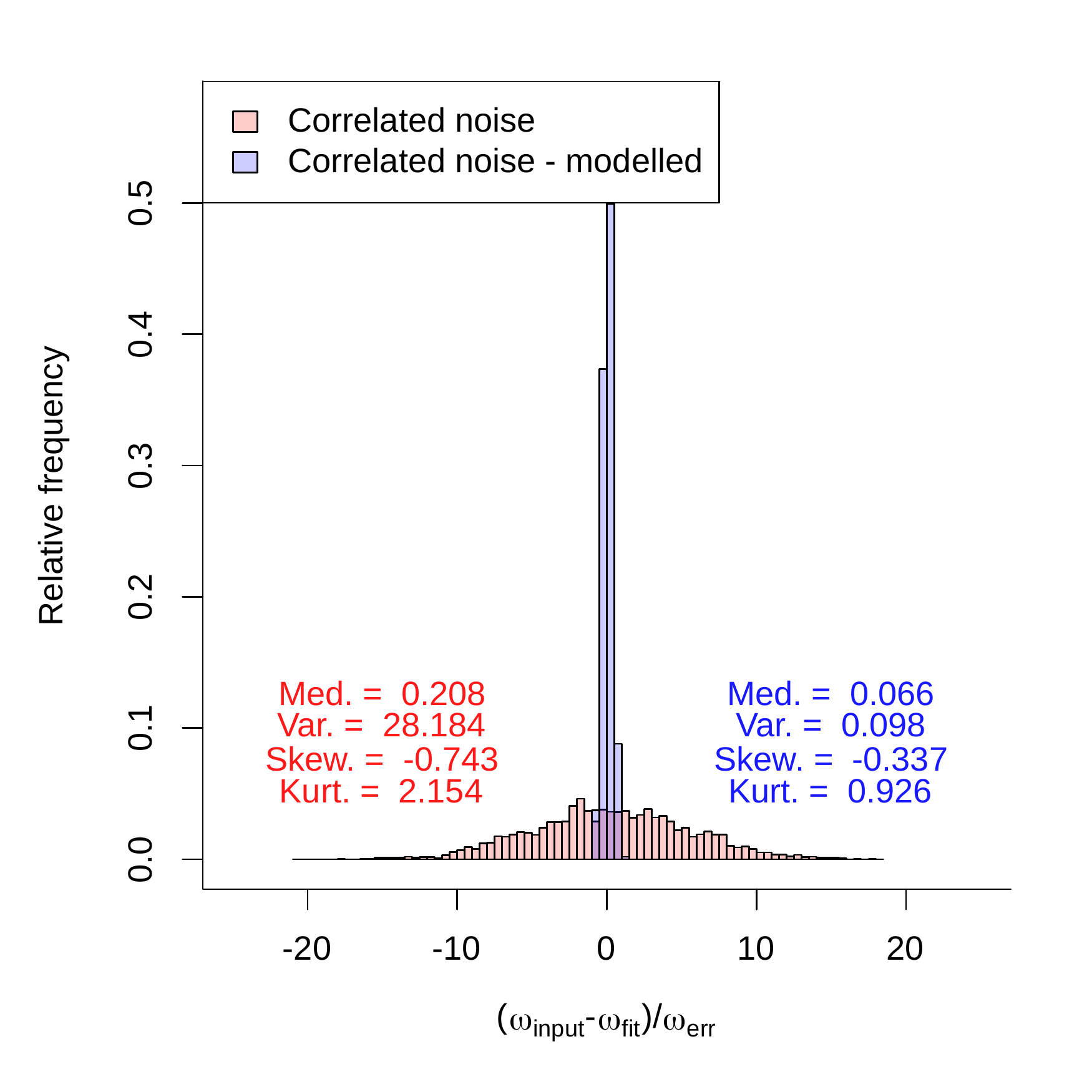}
    \captionof{figure}{Distribution of the differences between the input parameters and the fitted parameters for the same correlated noise model, without the noise fitting (red, i.e. same as on Fig. \ref{fig:rel_r_w}) and with the red noise handled through wavelet transformation (blue), scaled with the uncertainties estimated from the fitting. The median, variance, skewness, and kurtosis of the distributions is also shown.}
    \label{fig:TLCM_rel_r_w}
 \end{minipage}

\end{appendix}
\end{document}